\documentclass[journal]{IEEEtran}
\usepackage{tabularx}
\usepackage{color}
\usepackage{amsmath,amsfonts,amssymb,amscd,amsthm}
\usepackage{mathtools,mathrsfs}
\usepackage{mdwmath,mdwtab}
\usepackage{multirow}
\usepackage{cite}
\usepackage{booktabs}
\usepackage{enumitem}
\usepackage{siunitx}
\usepackage{nccmath}
\usepackage[english]{babel}
\setlocalecaption{english}{figure}{Fig.}
\setlocalecaption{english}{table}{TABLE}
\usepackage{nicefrac, xfrac}
\usepackage[compact]{titlesec}
\usepackage{nomencl}
\usepackage{booktabs, siunitx}
\usepackage{pifont}
\usepackage{twemojis}
\usepackage{threeparttable}
\usepackage{xcolor}
\usepackage{tikz}
\usepackage{graphicx}

\makeatother
\usepackage{utfsym}

\DeclareMathOperator{\diff}{d}
\newcommand{\ddt}{\tfrac{\diff}{\diff\!t}}
\newcommand{\norm}[1]{\left\Vert#1\right\Vert}

\definecolor{curveblue}{RGB}{0,114,189}
\definecolor{curveorange}{RGB}{217,83,25}
\definecolor{curveyellow}{RGB}{237,177,32}
\definecolor{curvepurple}{RGB}{126,47,142}

\definecolor{PVorange}{RGB}{237, 177, 32}
\definecolor{PVred}{RGB}{217, 83, 25}
\definecolor{PVblue}{RGB}{0, 114, 189}

\DeclareRobustCommand{\opencircle}[1]{%
	\tikz[baseline=-0.55ex]
	\draw[#1,line width=0.6pt] (0,0) circle[radius=0.55ex];%
}

\DeclareRobustCommand{\opentriangle}[1]{%
	\tikz[baseline=-0.45ex]
	\draw[#1,line width=0.6pt]
	(0,0.65ex) --
	(-0.65ex,-0.45ex) --
	(0.65ex,-0.45ex) -- cycle;%
}

\DeclareRobustCommand{\solidline}[1]{%
	\tikz[baseline=-0.55ex]
	\draw[#1, solid, line width=0.8pt]
	(0,0) -- (1.6em,0);%
}

\usepackage{xpatch}
\makeatletter
\xpatchcmd{\@thm}{\thm@headpunct{.}}{\thm@headpunct{}}{}{}
\makeatother

\begin{document}

\title{Constrained Grid-Forming Control for Photovoltaic Systems: A Control Barrier Function Approach}

\author{Amir~Afshari and
         Dominic~Gro\ss
 \thanks{}
 \thanks{A.~Afshari and D.~Gro\ss~are with the Department
 of Electrical and Computer Engineering, University of Wisconsin-Madison, Madison, WI 53706 USA. (e-mail: afshari2@wisc.edu and dominic.gross@wisc.edu)}}

% The paper headers

% make the title area
\maketitle

\IEEEpeerreviewmaketitle
\begin{abstract}
This paper leverages control barrier functions (CBFs) to systematically account for constraints of voltage source converters (VSCs) and photovoltaic (PV) resources in grid-forming (GFM) control. The proposed control combines (unconstrained) dual-port GFM control with control barrier functions (CBFs) to ensure reliable and safe operation of a GFM PV system under grid faults, sudden changes in irradiance, and load perturbations. First, CBFs are used to encode physical and operational limits of the PV system, namely current and voltage limits at its ac and dc terminals. Then, CBFs are used to restrict the dynamics of dual-port GFM control to within the VSC and PV limits during both transient and steady states. To this end, a cost function is defined that encodes trade-offs between forming the ac voltage phase angle/frequency and voltage magnitude under constraints. Then, real-time optimization is used to make the least modification, as measured by the cost function, to the (unconstrained) dual-port GFM control state to remain within the safe set encoded by the CBFs. Finally, the performance of the developed constrained GFM control is evaluated through EMT simulations in MATLAB/Simulink.
\end{abstract}


\section{Introduction}
Sustainability targets and low renewable-generation costs have driven widespread deployment of inverter-based resources (IBRs). Traditionally, IBRs use grid-following (GFL) control, which assumes a stable ac voltage and uses a phase-locked loop (PLL) for synchronization. This assumption becomes increasingly difficult to justify as the share of GFL IBRs increases~\cite{li2022revisiting}. To overcome limitations of GFL control, grid-forming (GFM) control is therefore envisioned to provide the stabilizing functions and voltage and frequency stiffness traditionally supplied by synchronous generators (SGs). Various GFM control methods have been developed, including droop control~\cite{chandorkar1993control}, virtual synchronous machines (VSMs), and dual-port control~\cite{subotic2024universal}. Apart from the provision of inertia, these controls exhibit broadly similar dynamic responses and share the core functions of imposing stable and self-synchronizing ac voltage dynamics~\cite[Sec.~3.4]{dorfler2023control}.

Although GFM IBRs are a significant step towards integrating renewable generation and energy storage at scale, they face challenges due to converter and resource limits. Among other aspects, converter current limits~\cite{baeckeland2024overcurrent} and modulation limits, and resource power limits~\cite{iraniparast2025networked} and dc voltage limits~\cite{chen2019power,afshari2024interactions} need to be considered. Notably, PV must be operated below the maximum power point (MPP) to maintain active-power reserves for frequency regulation. While the resulting curtailment entails foregone energy-market revenue, it may also yield revenue from ancillary-service provision. Moreover, large power deviations on the ac side can cause substantial dc-link voltage excursions and lead to dc-voltage instability in PV GFM systems~\cite{afshari2024interactions}. However, typical GFM controls do not explicitly account for all relevant constraints. When limits are considered, the focus is often restricted to one limit (e.g., current, power, or dc voltage).

% Cut for reasons of space, should be put back in a revision.
%
% Given these challenges, designing a reliable GFM control strategy requires addressing two main objectives: 1) withstanding grid contingencies and faults without damaging the converter or the primary energy source during transient and steady-state operation, and 2) providing grid-support functions that contribute to system stability under abnormal operating conditions.

% A wide range of studies has been devoted to developing control strategies that enable GFM converters to withstand abnormal grid conditions while maintaining stable operation. Nevertheless, most existing approaches assume that the dc-side voltage of a voltage source converter (VSC) remains adequately regulated regardless of weather or grid conditions and primarily focus on current-limiting functionality. 

Current-limiting strategies for GFM converters can be broadly classified into three categories: i) direct methods, which enforce current limits through modification of the switching signals or current-reference saturation; ii) indirect methods, which modify power or voltage references or shape the converter output impedance; and iii) hybrid methods, which combine multiple current-limiting mechanisms.

Switch-level current limiting~\cite[Sec.~III-A]{baeckeland2024overcurrent} suppresses overcurrent in a few switching cycles but effectively disables all upstream controls, often resulting in instability and reduced power quality. The widely used circular current limiter retains the current loop and reliably limits current without degrading power quality, but the converter remains prone to loss of synchronism~\cite[Sec.~V]{baeckeland2024overcurrent}. Other direct approaches, such as switching to current control with a fixed reference angle~\cite{zhang2009power}, exhibit similar drawbacks. The saturated current-reference angle affects both the trade-off between active and reactive current support and the converter's transient stability~\cite{rokrok2021transient}.

In contrast, the most widely used indirect method modifies the GFM voltage reference to emulate an increasing output impedance~\cite{paquette2014virtual}. This threshold virtual impedance significantly improves transient stability after short-circuit fault clearing but may fail to limit current during phase jumps~\cite{zeng2023hybrid} and active power overload~\cite[Sec.~VII]{GrossConstrained}. Its tuning is also sensitive to grid impedance~\cite{wu2021small}. While adaptive tuning can reduce this sensitivity~\cite{wu2021small} and hybrid threshold virtual impedance~\cite{zeng2023hybrid} can limit current during phase jumps, these refinements address individual limitations without ensuring reliable current limiting across all these conditions. Alternative indirect approaches retain voltage control by saturating outer-loop voltage magnitude and angle references~\cite{chen2020use} or limiting active/reactive power references~\cite{taul2019current}. These approaches improve dynamic response and transient stability, but limited control bandwidth can allow transient current limit violations. The approach in~\cite{GrossConstrained} removes inner control loops and projects the GFM controller onto predicted current constraints, requiring precise filter parameters and parallel active damping.

Hybrid methods combine the fast response of direct limiters with the transient stability of indirect limiters. For instance, a circular current limiter can be combined with anti-windup feedback to the inner voltage control and outer dVOC~\cite{desai2024saturation}. Although this improves post-fault transient stability, extensions may be required to maintain stability during active-power overloads. Similarly, combining current-reference limiting and dynamic virtual impedance enhances transient stability and voltage support under unbalanced faults~\cite{baeckeland2023novel}. These methods neglect resource and dc-bus dynamics and typically consider only VSC current limits.

% Regarding the PV systems controls, the study in \cite{neves2015unbalanced} developed a reactive power and voltage support control strategy for single-stage grid-connected PVs under unbalanced grid conditions through modifying the current reference.
% The authors in \cite{yazdani2019advanced} have proposed a proportional-resonant control augmented with inner voltage and current loops for a specific setup based on a GFM photovoltaic synchronous generator system in parallel with a GFL battery connected to an unbalanced infinite bus.
% The method presented in \cite{afshari2017control} is developed for controlling grid-connected two-stage PVs under unbalanced grid conditions and proposes a power reference modification method to limit the injected current during grid voltage drops.

For PV systems, current limiting also affects the dc voltage and PV power generation. For single-stage current-controlled PV systems, the control in~\cite{neves2015unbalanced} combines dc-link voltage regulation with current-reference modification to provide reactive power and voltage support under unbalanced grid conditions. For two-stage current-controlled PV systems, the control developed in~\cite{afshari2017control} coordinates power-reference modification and dc-side control to limit current and suppress dc-link voltage oscillations under unbalanced faults. Both approaches consider a current-controlled VSC tracking current references derived from power commands and the measured grid voltage. However, neither approach explicitly incorporates dc-voltage and modulation limits into the control design.

The authors in~\cite{yazdani2019advanced} consider VSM control of a GFM PV converter in parallel with a GFL battery converter. The VSM control is augmented with proportional-resonant controls for frequency and voltage that suppress active and reactive power oscillations under unbalanced grid conditions. Current limiting is attributed to an inner current controller, but the limiter implementation is not discussed.
%
%
% In \cite{gross2023towards}, a constrained GFM control is formulated as a one-step optimal problem with applications to PV systems, which is designed based on a steady-state model of single-stage PV systems and cannot guarantee the performance during transients.
%
The constrained GFM control in~\cite{gross2023towards} uses primal-dual dynamics to approximate the solution of a one-step optimal control problem incorporating GFM objectives and converter and PV constraints in a quasi-steady-state ac model, but does not guarantee transient constraint satisfaction and may be unstable under severe faults.

%  A constrained cooperative control for GFM renewables is proposed in \cite{afshariConstrained}, which considers steady-state constraints by selecting appropriate power setpoints and frequency droop gains, but it acts like a supervisory control and does not account for limits during transients.
%  A dc-voltage limiter control for GFM Pvs is used in \cite{chen2019power} to keep the dc voltage stable. The limiter is realized as a PI control that acts on the droop equation when dc voltage hits the limits.
%   In \cite{pawar2021grid}, the authors present a model-free GFM control strategy for two-stage PV systems operating under different modes. The proposed strategy adopts a current-reference saturation approach to handle short-circuit faults.
%   Transient stability and active protection of grids with a high share of GFM PVs are the subject of \cite{roy2022transient}. Moreover, impacts of irradiance on the synchronization of PVs are evaluated, and an active dc-link protection scheme for solar irradiance drops is proposed.

The constrained cooperative control for GFM units in~\cite{afshariConstrained} selects power setpoints and frequency droop gains to satisfy steady-state constraints but does not guarantee transient constraint satisfaction. The dc-voltage limiter for GFM PV systems in~\cite{chen2019power} uses PI feedback and derivative action to reduce the droop frequency command and mitigate dc undervoltage during overload but does not consider current limits. For two-stage PV systems, the model-free GFM control in~\cite{pawar2021grid} adapts its operating mode to the available power reserves and uses a modified current-reference saturation during short-circuit faults. An active dc-link protection scheme for GFM PV plants using circular current limiting is used in~\cite{roy2022transient} to protect against solar irradiance drops that can cause a loss of synchronism.

% Despite the significant progress reported in the literature, several challenges remain insufficiently addressed. 1) A substantial portion of the existing studies primarily focuses on current-limiting strategies for idealized GFM converters with stable dc-side dynamics. Consequently, critical physical constraints, such as converter modulation limits and primary resource limitations, e.g., dc-side voltage constraints in PVs, are often neglected. The few studies that consider PV models typically introduce modifications to existing control strategies to address either current limiting or dc-link voltage regulation.
% 2) Some designs involve a tradeoff between stability and performance that depends on grid conditions and converter parameters, which are not fully known in advance. Therefore, no single tuning works well for every fault type.
% 3) Most current-limiting methods are developed through modifications of existing control loops rather than through a systematic constraint-based design framework, which may cause performance issues. For instance, once a limiter is active, the converter partly or fully loses its voltage-source behavior, which can cause windup or loss of synchronization \cite{baeckeland2024overcurrent}.  As a result, guaranteeing constraint satisfaction while simultaneously achieving desirable grid-supporting performance remains a challenging task.

Three challenges remain: 1) A substantial share of studies primarily focus on ac current limiting under the assumption of constant dc voltage and neglect critical constraints such as modulation limits, dc voltage limits, and resource limits. The few studies that consider PV systems address either current limiting or dc-link voltage regulation. 2) Control tuning depends on grid conditions, converter parameters, and the type of contingency (e.g., a fault or overload), which are not known in advance. Moreover, no single tuning works well for all contingencies. 3) Most approaches add limiters to existing controls rather than explicitly incorporating constraints into the control design. Limiter activation can compromise voltage-source behavior and cause windup or loss of synchronization~\cite{baeckeland2024overcurrent}. Thus, simultaneously satisfying constraints and preserving the stabilizing features of GFM control remains challenging.

% To address these limitations, this paper proposes a constrained GFM control strategy for PV inverters using CBFs. First, operational and physical constraints of the GFM PV converter, such as dc-link voltage and ac current limits, are formalized. Scalar barrier functions are then established for individual constraints, where system safety is safety is formally guaranteed by ensuring the non-negativity of the CBFs. Subsequently, differentiating the CBFs along the system dynamics maps the state constraints onto the control input (i.e., converter ac voltage), yielding a safe set of admissible control actions. Finally, the constrained GFM control is formulated as an optimization problem subject to these CBF-derived safety constraints.

To address these limitations, this paper proposes a constrained GFM control strategy for PV converters using CBFs. First, scalar CBFs are formulated that encode operational and physical constraints of the GFM PV converter, such as dc-link voltage and ac current limits. Subsequently, differentiating the CBFs along the system dynamics maps the state constraints onto the control input (i.e., converter ac voltage), yielding a safe set of admissible control actions. A cost function is defined that encodes trade-offs between forming the ac voltage phase angle/frequency and voltage magnitude under constraints. Finally, the constrained GFM control is formulated as an optimization problem that makes the least modification, as measured by the cost function, to the (unconstrained) GFM control to remain within the safe set of admissible control actions. Notably, feasibility of the optimization problem formally ensures constraint satisfaction.

% \textcolor{red}{It is worth mentioning that although asymmetric faults are of considerable practical importance, this work focuses on symmetric three-phase faults. The primary objective is to establish a systematic constrained control framework for GFM PV systems before addressing the additional challenges introduced by unbalanced operation. To the authors' knowledge, constrained control strategies for GFM PVs that simultaneously enforce dc voltage, modulation, and current limits have received very limited attention. Furthermore, existing approaches largely rely on heuristic mechanisms or modifications of unconstrained controllers, rather than explicitly incorporating physical constraints. In contrast, the proposed framework provides a systematic and explicit means of embedding these constraints within the control synthesis. Accordingly, this work first establishes the proposed constrained control under symmetric fault conditions, following the development path observed in the broader literature on current-limiting control of ideal GFM converters. }

Although asymmetric faults are common in practice, this work focuses on symmetric three-phase faults. The primary objective is to develop a systematic framework for incorporating converter and PV constraints into GFM control that can be extended to unbalanced faults and operation. To the authors' knowledge, control strategies for GFM PV systems that simultaneously enforce dc voltage, modulation, and current limits have received very limited attention. Existing approaches largely rely on heuristic modifications of unconstrained controllers, rather than explicitly incorporating constraints into the control design. The proposed framework systematically incorporates these constraints into the control synthesis for the three-phase balanced case.

% In this regard, the main contributions of this study can be summarized as follows: 1) a systematic and scalable constrained GFM control framework for single-stage PV systems that explicitly incorporates converter and primary-source limits, e.g., output current, dc voltage, and modulation voltage, while allowing additional constraints to be included without redesigning the overall control structure, 2) a constrained GFM control that does not require knowledge of grid parameters  for its design and exhibits robustness against uncertainties in the converter parameters, and 3) a simplified control architecture that eliminates nested control loops (e.g., voltage and current loops) and additional parallel controls, thereby reducing tuning complexity and avoiding undesirable interactions among multiple control loops.

The main contributions of this work are: 1) a systematic and scalable constrained GFM control framework for single-stage PV systems that explicitly incorporates ac current, dc voltage, and modulation limits and accommodates additional constraints without redesigning the control structure; 2) a constrained GFM control that requires no exact knowledge of grid parameters and is robust to filter-parameter uncertainty; and 3) a simplified control architecture without nested voltage and current loops or additional parallel controls that reduces tuning complexity and eliminates the risk of adverse dynamic interactions among multiple nested control loops

\section{Preliminaries}
This section introduces preliminaries and reviews basic theoretical results on CBFs. 

\subsection{Mathematical Notation}
Given a vector $x \in \mathbb{R}^n$, where $\mathbb{R}$ denotes the set of real numbers, the Euclidean norm of $x$ is denoted by $\norm{x} \coloneqq \sqrt{x^{\top}x}$. Moreover, $\angle x$ denotes the angle of the 2D vector $x \in \mathbb{R}^2$. The Clarke transformation is denoted by
\begin{align}\label{eq.Clarke}
\mathbf{T}_{{\alpha \beta}} \coloneqq \frac{2}{3}\begin{bmatrix}
1 & -\frac{1}{2}& -\frac{1}{2} \\
 0 & \frac{\sqrt3}{2} & -\frac{\sqrt3}{2}%\\
 %\frac{1}{2}& \frac{1}{2}&\frac{1}{2}\\
\end{bmatrix}.
 \end{align}
Notably, assuming three-wire connection, any three-phase current or voltage $x_{\mathrm{abc}} \in \mathbb{R}^3$ can be equivalently expressed in $\alpha\beta$-frame as $x_{{\alpha \beta}} = \mathbf{T}_{{\alpha \beta}}x_{\mathrm{abc}} \in \mathbb{R}^2$. Moreover, the Park transform is used to represent sinusoidal signals in $\alpha\beta$-frame as constant signals in dq-frame. To this end, let ${\mathbf T}_{\mathrm{dq}}(\theta)\coloneqq \mathbf R(\theta)\mathbf T_{\alpha\beta}$ and 
$x_{\mathrm{dq}} = \mathbf{T}_{\mathrm{dq}}(\theta)x_{\mathrm{abc}} \in \mathbb{R}^2$ denote the Park transformation where $\theta \in [-\pi, \pi)$ is the (time-varying) reference angle and the 2D rotation matrix $\mathbf{R}(\theta)$ is defined as
\begin{align}\label{eq.R2D}
\mathbf{R}(\theta) \coloneqq \begin{bmatrix}
\cos (\theta) & \sin (\theta) \\
 -\sin (\theta) & \cos (\theta)
\end{bmatrix}.
 \end{align}

\subsection{Review of Control Barrier Functions}\label{sec.cbfConcept}
We briefly introduce CBFs and review their key properties, which will be used to develop a constrained GFM control for PV systems in the remainder. The interested reader is referred to \cite{ames2016control} for a comprehensive review of CBFs.

Consider a nonlinear control-affine system of the form
\begin{align}\label{eq.samplesys}
\ddt {x\left(t\right)} = f(x(t)) + g(x(t))u(t),
 \end{align}
where $x(t) \in \mathbb{R}^{n_x}$ and
$u(t) \in \mathbb{R}^{n_u}$ denote the state vector and the control input, and $\mathcal{U} \subseteq \mathbb{R}^{n_u}$ denotes the input constraints. The maps $f: \mathbb{R}^{n_x} \to \mathbb{R}^{n_x}$ and
$g: \mathbb{R}^{n_x} \to \mathbb{R}^{n_x \times n_u}$ are assumed to be locally Lipschitz \cite{khalil_nonlinear_2002}. For brevity, the explicit time dependence of $x$ and $u$ is dropped hereafter.

The goal of a CBF-based control is to certify the forward invariance of a safe set $\mathcal{C} \subset \mathbb{R}^{n_x}$, defined via a continuously differentiable function $h: \mathbb{R}^{n_x} \to \mathbb{R}$ as $\mathcal{C} \coloneqq \{ x \in \mathbb{R}^{n_x}: h(x) \geq 0 \}$. Accordingly, one can define the boundary and interior of $\mathcal{C}$ as $\partial\mathcal{C}\coloneqq \{ x \in \mathbb{R}^{n_x} : h(x) = 0  \}$ and $\mathrm{Int}(\mathcal{C})\coloneqq \{ x \in \mathbb{R}^{n_x} : h(x) > 0  \}$. The function $h$ is a CBF for \eqref{eq.samplesys} if there exists a constant $\alpha \in \mathbb{R}_{>0}$ such that, for every $x \in \mathbb{R}^{n_x}$, there exists $u \in \mathcal{U}$ satisfying
\begin{align}\label{eq.cbfDef}
\ddt h(x) = \frac{\partial h}{\partial x}\left(f(x)+g(x)u\right) \geq -\alpha h(x),
\end{align}
i.e., $h(x)\geq0$ will remain nonnegative. The system \eqref{eq.samplesys} is said to be safe if $x(0) \in \mathcal{C}$ implies $x(t) \in \mathcal{C}$ for all $t \geq 0$, i.e., $\mathcal{C}$ is forward invariant. Intuitively, $h(x)$ acts as a scalar safety margin. A positive value indicates that the state lies in the interior of $\mathcal{C}$, zero corresponds to its boundary, and a negative value signals a constraint violation. As the trajectory approaches $\partial\mathcal{C}$, the condition \eqref{eq.cbfDef} compels the control input to counteract any outward drift, thereby keeping the system within the safe set. 
%The parameter $\alpha$ governs the aggressiveness of this corrective action, i.e., for small $\alpha$ the system is approach the boundary more slowly while large $\alpha$ result in a more abrupt limiting action close to the constraint. 
The parameter $\alpha$ governs the limiting response, with small $\alpha$ slowing the approach to the boundary and large $\alpha$ producing more abrupt action near the constraint. In practice, a control input $u$ satisfying \eqref{eq.cbfDef} is typically obtained by solving a constrained optimization problem.

\section{Single-Stage PV System Model \& Constraints}
\begin{figure}[b!]
  \centerline{\includegraphics[scale=0.23,trim={10cm 10cm 10cm 12cm}]{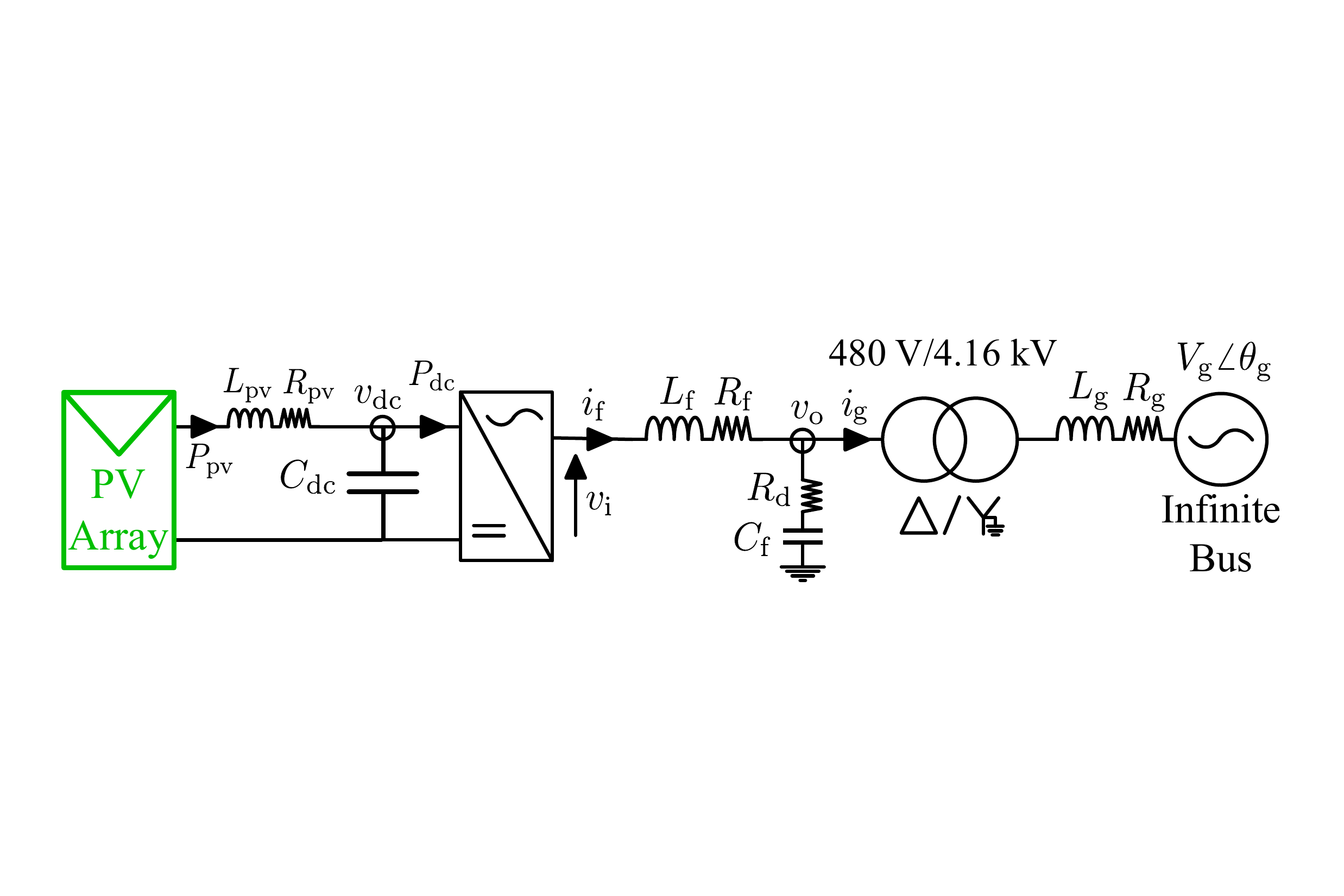}}
  \caption{Single-stage  GFM PV unit connected to an infinite bus.}
  \label{fig.PVsingleBus}
  \end{figure}

The single-stage PV system in Fig.~\ref{fig.PVsingleBus} consists of a photovoltaic array, a dc-link capacitor that serves as an intermediate energy buffer, and a two-level VSC connected to the grid through an LC filter and step-up transformer. 
This section first introduces the dynamics as well as physical and operational constraints of the single-stage PV system. Subsequently, CBFs encoding the constraints are developed.

\subsection{Model of the Photovoltaic Array}\label{subsec:PVarray}
 \begin{figure}[ht!!!]
  \centerline{\includegraphics[scale=0.33,trim={7cm 0.1cm 7cm 0.1cm}]{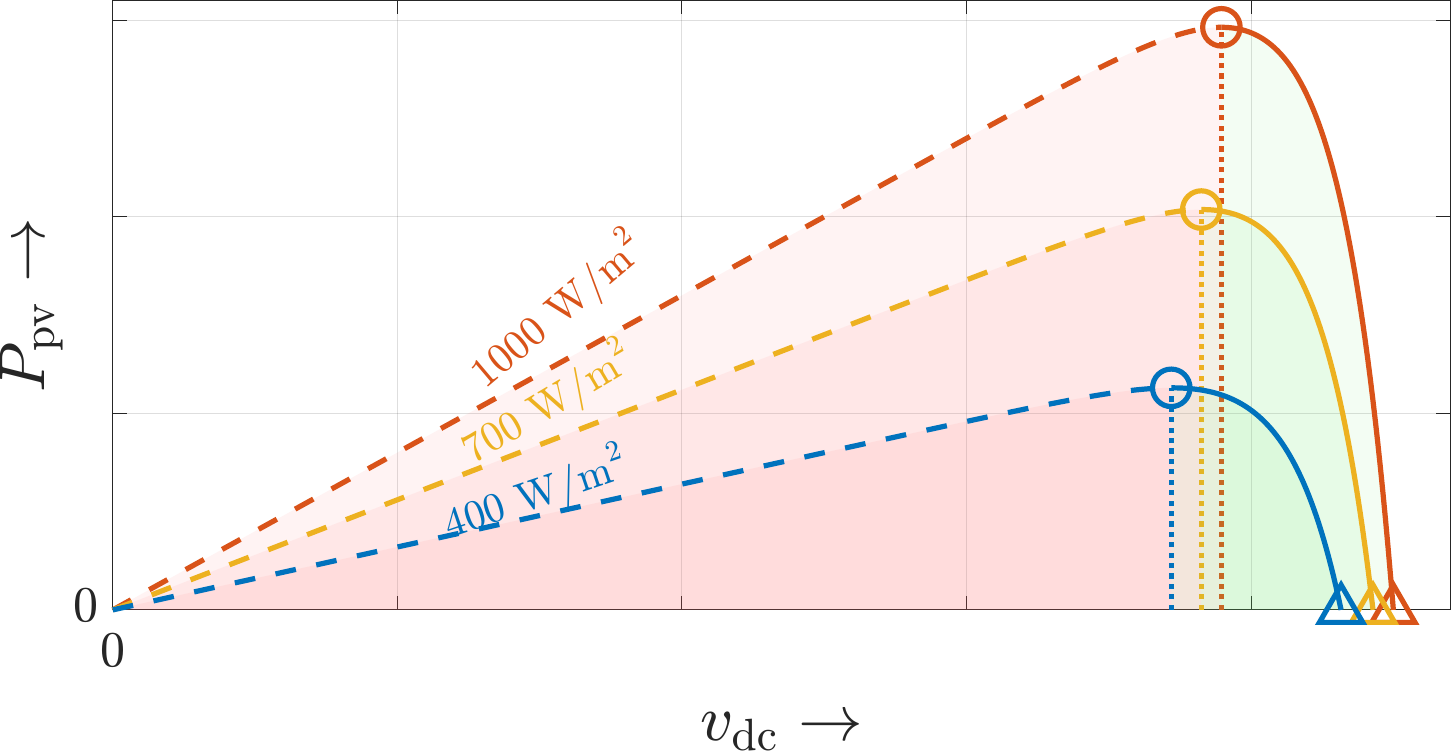}}
\caption{$P_{\mathrm{pv}}-v_{\mathrm{dc}}$ characteristics for irradiance levels $400~\mathrm{W/m^2}$ (\solidline{PVblue}), $700~\mathrm{W/m^2}$ (\solidline{PVorange}), and $1000~\mathrm{W/m^2}$ (\solidline{PVred}) with corresponding MPPs (\opencircle{PVblue}), (\opencircle{PVorange}), and (\opencircle{PVred}), open-circuit voltages $v_{\mathrm{dc}}^{\mathrm{oc}}$ (\opentriangle{PVblue}), (\opentriangle{PVorange}), and (\opentriangle{PVred}), and unstable ({\setlength{\fboxsep}{1pt}\colorbox[rgb]{1,0.8627,0.8627}{\rule{0pt}{1.2ex}\hspace{0.9em}}}) and stable ({\setlength{\fboxsep}{1pt}\colorbox[rgb]{0.8627,0.9765,0.8627}{\rule{0pt}{1.2ex}\hspace{0.9em}}}) regions.}
  \label{fig.curve}
  \end{figure}
  For a given temperature, the nonlinear characteristics of a PV array can be modeled by the nonlinear $P_{\mathrm{pv}}-v_{\mathrm{dc}}$ characteristic shown in Fig.~\ref{fig.curve}. The maximum admissible dc voltage for a PV system is $v_{\mathrm{dc}}^{\mathrm{oc}} \in \mathbb{R}_{>0}$, i.e., the open-circuit voltage, which corresponds to zero power. On the other hand, the minimum dc voltage is $v_{\mathrm{dc}}^{\text{mpp}} \in \mathbb{R}_{>0}$, which corresponds to the MPP. Moreover, the aggregate collector network dynamics can be approximately modeled by
   \begin{align}\label{eq.pvpanel}
  	\tau_{\mathrm{pv}}\ddt P_{\mathrm{dc}}
  	= -P_{\mathrm{dc}}+P_{\mathrm{pv}},
  \end{align}
  where $\tau_{\mathrm{pv}} \coloneqq L_{\mathrm{pv}}/R_{\mathrm{pv}}$ is the collector-network time constant, and $L_{\mathrm{pv}} \in \mathbb{R}_{>0}$ and $R_{\mathrm{pv}} \in \mathbb{R}_{>0}$ denote the aggregate collector-network inductance and resistance, respectively.

  Next, we define the sensitivity of the $P_{\mathrm{pv}}-v_{\mathrm{dc}}$ characteristic at a dc voltage ${v_{\mathrm{dc}}^{\star}}\in \mathbb{R}_{>0}$ as $k_{\mathrm{pv}} \coloneqq -\partial P_{\mathrm{pv}}/\partial v_{\mathrm{dc}}|_{v_{\mathrm{dc}}=v_{\mathrm{dc}}^{\star}}$. 
  The stable operating region is defined as the part of the $P_{\mathrm{pv}}-v_{\mathrm{dc}}$ characteristic for which $k_{\mathrm{pv}} \ge 0$ such that a decrease in dc voltage results in increased power generation. On the other hand, if the dc voltage drops below $v_{\mathrm{dc}}^{\text{mpp}}$, the PV system enters the unstable operating region, in which a decrease in dc voltage results in decreased power generation. Moreover, exceeding $v_{\mathrm{dc}}^{\mathrm{oc}}$ may cause premature dc-link capacitor failures and reverse current that may result in severe damage through PV cell overheating. Overall, the dc voltage needs to satisfy $v_{\mathrm{dc}}^{\text{mpp}} \le v_{\mathrm{dc}} \le v_{\mathrm{dc}}^{\mathrm{oc}}$ to ensure both stable and safe operation. To provide GFM functions, $v_{\mathrm{dc}}^{\text{mpp}} < v_{\mathrm{dc}} \le v_{\mathrm{dc}}^{\mathrm{oc}}$ is used to maintain an active-power reserve for frequency control.

%In this study, we focus on single-stage PVs, which, due to their reduced degrees of freedom, are more challenging to control. 

\subsection{Converter Modeling}\label{subsec:convmodel}
The GFM controller generates the ac-voltage control input $v_{\mathrm{i,abc}}\in \mathbb{R}^3$ for the VSC. The maximum allowable voltage magnitude for the line-to-neutral peak voltage of a two-level three-phase converter is $V^{\max}_{\mathrm{mod}} \coloneqq \frac{1}{2}v_{\mathrm{dc}}$. The ac filter inductance, resistance, and capacitance are denoted by $L_{\mathrm{f}}\in\mathbb{R}{>0}$, $R_{\mathrm{f}}\in\mathbb{R}{>0}$, and $C_{\mathrm{f}}\in\mathbb{R}_{>0}$. For brevity of the presentation, we assume a passive damping resistor $R_{\mathrm{d}}\in\mathbb{R}_{>0}$ is used. To reduce losses a virtual damping resistor can be used. The dynamics of the filter current $i_{\mathrm{f,abc}}\in \mathbb{R}^3$ can be modeled in $\text{dq}$ coordinates as
\begin{subequations}\label{eq.idq}
	\begin{align}
\ddt {i}_{\mathrm{fd}} = \frac{-R_\mathrm{f}}{L_\mathrm{f}}i_{\mathrm{fd}}+ \omega^{\star}i_{\mathrm{fq}}+\frac{1}{L_\mathrm{f}}\left( v_{\mathrm{id}}-v_{\mathrm{od}}\right), \label{eq.id} \\
\ddt {i}_{\mathrm{fq}} = \frac{-R_\mathrm{f}}{L_\mathrm{f}}i_{\mathrm{fq}} - \omega^{\star}i_{\mathrm{fd}}+\frac{1}{L_\mathrm{f}}\left( v_{\mathrm{iq}} - v_{\mathrm{oq}} \right).\label{eq.iq}
\end{align}
\end{subequations}
Here, $v_{\mathrm{od}} = v_{\mathrm{fd}} + R_{\mathrm{d}}(i_{\mathrm{fd}}-i_{\mathrm{gd}})$ and $v_{\mathrm{oq}} = v_{\mathrm{fq}} + R_{\mathrm{d}}(i_{\mathrm{fq}}-i_{\mathrm{gq}})$ include the voltage drop across the damping resistor and $i_{\mathrm{g,dq}}$ denotes current flowing from the filter into the grid.

Notably, the semiconductor temperature is subject to strict limits to avoid damage. The standard approach to protect semiconductor switches against overheating is to limit the current magnitude $\norm{i_{\mathrm{f}}}$ to the long-term current limit $I_{\max} \in \mathbb{R}_{>0}$. The long-term limit $I_{\max}$ is typically conservative relative to short-duration pulse ratings. In other words, brief overcurrents are tolerable while sustained ones are not. This principle is analogous to inverse-time overcurrent protection~\cite{IEEEstd} that defines trip-time characteristics that vary inversely with the current magnitude.

% The dynamics of the filter capacitor voltage $v_{\mathrm{f,dq}}$ are given by
% % \begin{subequations}\label{eq.vfdq}
% % \begin{align}
% % \ddt v_{\mathrm{fd}} &= \frac{i_{\mathrm{fd}}-i_{\mathrm{gd}}}{C_{\mathrm{f}}}
% % + \omega^{\star}v_{\mathrm{fq}}, \label{eq.vfd}\\
% % \ddt v_{\mathrm{fq}} &= \frac{i_{\mathrm{fq}}-i_{\mathrm{gq}}}{C_{\mathrm{f}}}
% % - \omega^{\star}v_{\mathrm{fd}}. \label{eq.vfq}
% % \end{align}
% % \end{subequations}
% %
% \begin{align*}
% \ddt v_{\mathrm{fd}} = \frac{i_{\mathrm{fd}}-i_{\mathrm{gd}}}{C_{\mathrm{f}}} + \omega^{\star}v_{\mathrm{fq}}, \quad \ddt v_{\mathrm{fq}} = \frac{i_{\mathrm{fq}}-i_{\mathrm{gq}}}{C_{\mathrm{f}}} - \omega^{\star}v_{\mathrm{fd}}.
% \end{align*}
% %
% Finally, $i_{\mathrm{g,dq}}$ denotes current flowing from the filter into the grid and the output voltage $v_{\mathrm{od}} = v_{\mathrm{fd}} + R_{\mathrm{d}}(i_{\mathrm{fd}}-i_{\mathrm{gd}})$ and $v_{\mathrm{oq}} = v_{\mathrm{fq}} + R_{\mathrm{d}}(i_{\mathrm{fq}}-i_{\mathrm{gq}})$ includes the voltage drop across the damping resistor.
% %
% % \begin{subequations}\label{eq.vodq}
% % \begin{align}
% % v_{\mathrm{od}} &= v_{\mathrm{fd}} + R_{\mathrm{d}}(i_{\mathrm{fd}}-i_{\mathrm{gd}}), \label{eq.vod}\\
% % v_{\mathrm{oq}} &= v_{\mathrm{fq}} + R_{\mathrm{d}}(i_{\mathrm{fq}}-i_{\mathrm{gq}}). \label{eq.voq}
% % \end{align}
% % \end{subequations}
% %
% Notably, $v_{\mathrm{dc}}$, $i_{\mathrm{f,dq}}$, $v_{\mathrm{o,dq}}$, and $P_{\mathrm{dc}}$ are measured and available for control.

Finally, let $C_{\mathrm{dc}} \in \mathbb{R}_{>0}$ and $P_{\mathrm{ac}}\in \mathbb{R}$ denote the dc-side capacitance and VSC ac-side power injection. Using peak voltage and current bases, the ac power is $P_{\mathrm{ac}}\coloneqq v_{\mathrm{i,dq}}^{\top}i_{\mathrm{f,dq}}$ in per unit. Assuming a lossless converter, the dynamics of the energy $\tfrac{1}{2}C_{\mathrm{dc}} v_{\mathrm{dc}}^2$ stored in the dc-link capacitor can be modeled by the energy-balance model
\begin{align}\label{eq.dccap}
\ddt \tfrac{1}{2}C_{\mathrm{dc}} v_{\mathrm{dc}}^2 = P_{\mathrm{dc}} - P_{\mathrm{ac}}.
\end{align}

A crucial quantity is the energy relation $\tau_{\mathrm{e}}S_{\mathrm{b}} = \frac{1}{2}C_{\mathrm{dc}} v_{\mathrm{dc,b}}^2$, where $\tau_{\mathrm{e}}\in \mathbb{R}_{>0}$, $S_{\mathrm{b}}\in \mathbb{R}_{>0}$, and $v_{\mathrm{dc,b}}\in \mathbb{R}_{>0}$ denote the dc-link energy storage in seconds, the base power, and the base dc voltage, respectively.

\section{Review of Unconstrained Dual-Port GFM Control for Single-Stage PV Systems}
Conventional GFM controls, such as droop control, virtual synchronous machine (VSM) control, and dVOC, are primarily developed for grid voltages within the normal operating range. Moreover, although widely adopted in the literature, these controls generally require a tightly controlled dc voltage. This assumption does not hold in practice for renewable sources such as PV. In that case, neglecting dc bus dynamics in the design of GFM controls can lead to instability and degraded performance under load fluctuations or changes in weather conditions even when the grid voltage remains within the normal operating range.

% For the unconstrained GFM control design, we linearize the PV characteristic at the nominal operating point $(v_{\mathrm{dc}}^{\star},P_{\mathrm{pv}}^{\star})$ as
%%
%% \begin{align}\label{eq.pvpanel}
%	% \left(\tau_{\text{pv}}s+1\right) \left( P_{\text{pv}}\left(s\right) - P_{\text{pv}}^{\star}\left(s\right) \right) = - k_{\text{pv}}\left( v_{\mathrm{dc}}\left(s\right) - v_{\mathrm{dc}}^{\star}\left(s\right) \right),
%	% \end{align}
%% 
%\begin{align}\label{eq.pvpanel}
%	\tau_{\mathrm{pv}}\ddt P_{\delta,\mathrm{dc}}
%	= -P_{\delta,\mathrm{dc}}-k_{\mathrm{pv}}v_{\delta,\mathrm{dc}},
%\end{align}
%%
%where $P_{\delta,\mathrm{dc}}\coloneqq P_{\mathrm{dc}}-P_{\mathrm{pv}}^{\star}$ and $v_{\delta,\mathrm{dc}}\coloneqq v_{\mathrm{dc}}-v_{\mathrm{dc}}^{\star}$ are the deviations from the nominal operating point. 

To account for dc bus dynamics~\eqref{eq.dccap}, this study builds on dual-port GFM control that simultaneously regulates the PV dc bus voltage and the ac active power/frequency~\cite{subotic2024universal,afshariConstrained}. Unlike conventional GFM controls that map deviations of the active power from its reference to frequency deviations without accounting for the dc bus voltage, dual-port GFM control regulates the frequency as a function of the deviation of the dc bus voltage from its reference. Specifically, we obtain a reference frequency 
\begin{align}\label{eq.dualPort1}
\omega^{\text{dp}}\left(s\right) = \omega^{\star} +\frac{{k_\mathrm{p}}{\left(\tau_{\mathrm{p}} s + k_{\mathrm{pv}}\right)}}{{\left( {{{\tau}_{\mathrm{pv}}}s + 1} \right)}}v_{{\delta},\mathrm{dc}}\left(s\right),
\end{align}
using the (unconstrained) controller~\cite{afshariConstrained} with proportional gain ${k_\mathrm{p}} \in \mathbb{R}_{>0}$, derivative gain $\tau_{\mathrm{p}}  \in \mathbb{R}_{>0}$, time constant $\tau_{\mathrm{pv}}\in\mathbb{R}_{>0}$ of the  PV model~\eqref{eq.pvpanel}, and deviation $v_{{\delta},\mathrm{dc}} \eqqcolon v_{\mathrm{dc}} - v_{\mathrm{dc}}^{\star}$ of the dc-bus voltage from its reference $v_{\mathrm{dc}}^{\star} \in \mathbb{R}_{>0}$. Notably, the steady-state frequency droop constant is equal to ${k_\mathrm{p}}$. Moreover, the effective time constant $\tau_{\mathrm{pv}}\in\mathbb{R}_{\ge 0}$ approximately models cable dynamics and is only used to tune the unconstrained controller to meet small-signal stability specifications (see \cite[Subsection IV-B]{afshariConstrained} for more details).

Moreover, considering the voltage setpoint $V^{\star}\in \mathbb{R}_{>0}$ and voltage droop coefficient $m_{\mathrm{q}}\in \mathbb{R}_{>0}$, the ac voltage magnitude is controlled through the standard reactive power-voltage droop control $V_{\mathrm{ref}}(s) = \frac{1}{\tau_{\mathrm{v}}s+1 }(V^{\star} + m_{\mathrm{q}} ( Q^{\star} - \tilde{Q}(s)))$, where $\tau_{\mathrm{v}} \in \mathbb{R}_{>0}$, $V_{\mathrm{ref}} \in \mathbb{R}$, $ Q^{\star} \in \mathbb{R}$, and $\tilde{Q} \in \mathbb{R}$ are the voltage control time constant, voltage reference magnitude, reactive power setpoint, and filtered reactive power, respectively. Here, $\tilde{Q}$ is the filtered value of the per-unit reactive power $Q\coloneqq v_{\mathrm{oq}}i_{\mathrm{gd}}-v_{\mathrm{od}}i_{\mathrm{gq}}$ at the VSC filter capacitor. Overall, the unconstrained GFM voltage reference is given by $v_{\mathrm{i,dq}}^{\text{ref}} = \left( V_{\mathrm{ref}}, 0 \right)$ and ${\theta}^{\text{dp}}$.

Compared to other state-of-the-art GFM controls, dual-port GFM control \eqref{eq.dualPort1} controls the dc-bus voltage to within acceptable limits in steady state after moderate load perturbations and changes in solar irradiance~\cite{afshariConstrained}. However, because voltage and current limits are not explicitly accounted for in \eqref{eq.dualPort1}, it may not be able to maintain operation within the PV system constraints during transients, severe contingencies (e.g., faults), large load perturbations, and overloads. Therefore, in the following, we will use the unconstrained control to generate a candidate GFM voltage reference. If that candidate voltage results in a violation of converter constraints, the least modification (as measured by a suitable cost function) to the unconstrained controller dynamics will be made to ensure operation of the PV system within its constraints.

\section{Constrained GFM Control for Photovoltaics}

In this section, the proposed constrained GFM control is presented, together with its main properties and the algorithm used for its implementation. The CBF-based safe sets introduced previously are employed as the constraint sets of the proposed control strategy.

\subsection{Control Barrier Functions for Single-Stage PV Systems}
In this section, individual CBFs are constructed for each of the constraints discussed in the previous section. Notably, the resulting CBF constraints~\eqref{eq.cbfDef} can ultimately be expressed as bounds on the  GFM reference for the ac voltage modulated by the converter. 

\subsubsection{AC Current Limit}
Considering the stored inductor energy $E_{\mathrm{f}}$ and its maximum $E^{\max}_{\mathrm{f}}$ at the current limit, we define the CBF 
\begin{align}\label{eq.Hi}
h_{\mathrm{i}}^{\max} \coloneqq \underbrace{\tfrac{1}{2} L_{\mathrm{f}} I_{\max} ^2}_{\eqqcolon E^{\max}_{\mathrm{f}}} - \underbrace{\tfrac{1}{2} L_{\mathrm{f}} \norm{i_{\mathrm{f,dq}}}^2}_{\eqqcolon E_{\mathrm{f}}}
\end{align}
associated with the ac current limit.
As discussed in Sec.~\ref{sec.cbfConcept}, the current limit can be enforced by selecting the ac voltage $v_{\mathrm{i,dq}} \in \mathbb{R}^2$ such that the corresponding CBF condition~\eqref{eq.cbfDef} is satisfied. Taking the time derivative of \eqref{eq.Hi} along the current dynamics \eqref{eq.idq} results in
\begin{align}\label{eq.HiDeri}
\ddt h_{\mathrm{i}}^{\max} =  \underbrace{i_{\mathrm{f,dq}}^{\top}v_{\mathrm{o,dq}}}_{\coloneqq P_{\mathrm{out}}}  - \underbrace{i_{\mathrm{f,dq}}^{\top}v_{\mathrm{i,dq}}}_{\coloneqq P_{\mathrm{ac}}} + \underbrace{R_{\mathrm{f}} \norm{i_{\mathrm{f, dq}}}^2}_{\coloneqq P_{\mathrm{loss}}} .
 \end{align}
 
 Using \eqref{eq.cbfDef}, the set of ac voltages that are admissible with respect to the ac current limit is given by
\begin{align*}%\label{eq.HiSet}
    \mathcal{C}^{\max}_{\mathrm{i}} \!\coloneqq\! \left\{ v_{\mathrm{i,dq}} \!\in \mathbb{R}^{2} \,|\, i_{\mathrm{f,dq}}^{\top}v_{\mathrm{i,dq}} \!\le\! \alpha_{\mathrm{i}}^{\max} h_{\mathrm{i}}^{\max} \!+\!P_{\mathrm{out}}\!+\!P_{\mathrm{loss}}\right\}\!,
\end{align*}
where $\alpha_{\mathrm{i}}^{\max}\in \mathbb{R}_{>0}$ is a design parameter that determines how far away from the current limit and how aggressively the controller will start to slow the increase in current magnitude. This influences both the current rise time during fault inception and how fast the current returns to the safe set if a temporary constraint violation occurs.

Notably, the rate of change of the ac filter inductor energy is $\ddt E_{\mathrm{f}} = P_\mathrm{ac}-P_{\mathrm{out}}-P_{\mathrm{loss}}$. Rearranging the inequality defining $\mathcal{C}^{\max}_{\mathrm{i}}$ results in $\ddt E_{\mathrm{f}} \leq \alpha_{\mathrm{i}}^{\max} h_{\mathrm{i}}^{\max}$. In other words, the increase of energy stored in the filter inductor is limited as $\norm{i_{\mathrm{f,dq}}} \to I_{\max} $ (i.e., $h_{\mathrm{i}}^{\max} \to 0$). When $\norm{i_{\mathrm{f,dq}}} = I_{\max} $, the active power $P_{\mathrm{ac}}$ flowing into the filter inductor from the VSC switches has to be less than or equal to the power $P_{\mathrm{out}}+P_{\mathrm{loss}}$ that is dissipated in and flowing out of the filter inductor towards the grid. This ensures that the inductor energy and current cannot increase further. We emphasize that larger $\alpha$ allows for a faster increase in inductor energy $E_{\mathrm{f}}$ at any given ac current magnitude which will in turn result in more aggressive limiter action close to the current limit. In contrast, smaller $\alpha$ results in a slower approach to the current limit.

\subsubsection{DC Voltage Limits}
Compared to a VSC with battery energy storage, the primary challenge of implementing GFM control for a single-stage PV system is the need to account for dc-side voltage dynamics and limits. As discussed in Sec.~\ref{subsec:PVarray}, the dc voltage needs to satisfy $v_{\mathrm{dc}}^{\text{mpp}} \le v_{\mathrm{dc}} \le v_{\mathrm{dc}}^{\mathrm{oc}}$. To this end, we define the normalized dc-bus energy CBFs as
%
% \begin{subequations}
% \begin{align}\label{eq.He}
% h_{\mathrm{dc}}^{\max} &\coloneqq \tfrac{1}{2} C_{\mathrm{dc}}  \frac{{v_{\mathrm{dc}}^{\mathrm{oc}}}^2 -  v_{\mathrm{dc}}^2}{\tau_{\mathrm{e}}S_{\mathrm{b}}},\\
%  h_{\mathrm{dc}}^{\min} &\coloneqq \tfrac{1}{2} C_{\mathrm{dc}}  \frac{v_{\mathrm{dc}}^2 -  {v_{\mathrm{dc}}^{\text{mpp}}}^2}{\tau_{\mathrm{e}}S_{\mathrm{b}}}. 
% \end{align}
% \end{subequations}
\begin{align}\label{eq.He}
h_{\mathrm{dc}}^{\max} \coloneqq \tfrac{1}{2} C_{\mathrm{dc}}  \frac{{v_{\mathrm{dc}}^{\mathrm{oc}}}^2 -  v_{\mathrm{dc}}^2}{\tau_{\mathrm{e}}S_{\mathrm{b}}}, \;
 h_{\mathrm{dc}}^{\min} \coloneqq \tfrac{1}{2} C_{\mathrm{dc}}  \frac{v_{\mathrm{dc}}^2 -  {v_{\mathrm{dc}}^{\text{mpp}}}^2}{\tau_{\mathrm{e}}S_{\mathrm{b}}}. 
\end{align}
Note that to ease the interpretation and control tuning, a normalization by $\tau_{\mathrm{e}} S_{\mathrm{b}}$ is introduced to express the CBFs in per unit energy. Enforcing the conditions $h_{\mathrm{dc}}^{\max} \ge 0$ and $h_{\mathrm{dc}}^{\min} \ge 0$ ensures that the dc-link voltage remains within the safe operating region $v_{\mathrm{dc}}^{\text{mpp}} \le v_{\mathrm{dc}} \le v_{\mathrm{dc}}^{\mathrm{oc}}$. The time derivatives of the CBFs \eqref{eq.He} along the dc-bus energy dynamics \eqref{eq.dccap} are 
%
% \begin{subequations}
% \begin{align}\label{eq.HeDot}
% \ddt h_{\mathrm{dc}}^{\max} &=   \frac{P_{\mathrm{ac}} - P_{\mathrm{dc}} }{\tau_{\mathrm{e}}S_{\mathrm{b}}},\\
%  \ddt h_{\mathrm{dc}}^{\min} &=  \frac{P_{\mathrm{dc}} - P_{\mathrm{ac}}}{\tau_{\mathrm{e}}S_{\mathrm{b}}}.
% \end{align}
% \end{subequations}
%
\begin{align}\label{eq.HeDot}
\ddt h_{\mathrm{dc}}^{\max} =   \frac{P_{\mathrm{ac}} - P_{\mathrm{dc}} }{\tau_{\mathrm{e}}S_{\mathrm{b}}},\quad  \ddt h_{\mathrm{dc}}^{\min} =  \frac{P_{\mathrm{dc}} - P_{\mathrm{ac}}}{\tau_{\mathrm{e}}S_{\mathrm{b}}}.
\end{align}
Thus, the sets of ac voltages that are admissible with respect to the dc-bus voltage limits are
\begin{subequations}
	\begin{align}\label{eq.HeConstrai}
\!\mathcal{C}^{\max}_{\mathrm{dc}} &\!\coloneqq\! \left\{ v_{\mathrm{i,dq}} \in \mathbb{R}^{2} | {i_{\mathrm{f,dq}}^{\top}v_{\mathrm{i,dq}} \ge P_{\mathrm{dc}}-\alpha_{\mathrm{dc}}^{\max} h_{\mathrm{dc}}^{\max}}\right\}\!,\\
\!\mathcal{C}^{\min}_{\mathrm{dc}} &\!\coloneqq\! \left\{ v_{\mathrm{i,dq}} \in \mathbb{R}^{2} | i_{\mathrm{f,dq}}^{\top}v_{\mathrm{i,dq}}\le \alpha_{\mathrm{dc}}^{\min} h_{\mathrm{dc}}^{\min} \!+\!P_{\mathrm{dc}}\right\}\!,
\end{align}
\end{subequations}

where $\alpha_{\mathrm{dc}}^{\max}\in \mathbb{R}_{>0}$ and $\alpha_{\mathrm{dc}}^{\min}\in \mathbb{R}_{>0}$ are design constants, and it is assumed that $P_{\mathrm{dc}}$ is measured and can be used to construct the constraint sets $\mathcal{C}^{\max}_{\mathrm{dc}}$ and $\mathcal{C}^{\min}_{\mathrm{dc}}$.

\subsubsection{Modulation Limit}
As discussed in Sec.~\ref{subsec:convmodel}, the maximum line-to-neutral peak voltage a two-level VSC can synthesize is $V^{\max}_{\mathrm{mod}} \coloneqq \frac{1}{2}v_{\mathrm{dc}}$, i.e., $V^{\max}_{\mathrm{mod}}$ changes depending on the PV array operating point. Notably, this is a state-dependent input constraint, i.e., no CBF is constructed. Consequently, the magnitude $\Vert v^{\mathrm{rms}}_{\mathrm{i,dq}}\Vert$ of the line-to-line root-mean-square (RMS) voltage is limited to
\begin{align}\label{eq.modConstra}
    \mathcal{C}_{\mathrm{mod}} &\coloneqq \left\{ v_{\mathrm{i,dq}} \in \mathbb{R}^{2} \;|\; \norm{v^{\mathrm{rms}}_{\mathrm{i,dq}}} \le {\frac{\sqrt{3}}{\sqrt{2}}} \frac{1}{2}v_{\mathrm{dc}} \right\}.
\end{align}

\subsection{Constrained GFM Control}
%  The novel constrained GFM control is obtained by limiting the dynamics of the unconstrained dual-port GFM control \eqref{eq.dualPort2} and \eqref{eq.Vdroop} to the safe set ${\mathcal{C}}^{\text{pv}} \coloneqq {\mathcal{C}}^{\max}_{\mathrm{i}} \cap \mathcal{C}_{\mathrm{mod}} \cap \mathcal{C}_{\mathrm{dc}}^{\max} \cap \mathcal{C}_{\mathrm{dc}}^{\min}$ obtained as the intersection of the CBFs. Specifically, if the (unconstrained) reference voltage ${\theta}^{\text{dp}}$ and $V_{\mathrm{ref}}$ obtained from \eqref{eq.dualPort2} and \eqref{eq.Vdroop} lies within the intersection of the CBF-based safe sets, i.e., ${\mathcal{C}}^{\text{pv}}$ all system constraints are satisfied and ${\theta}^{\text{dp}}$ and $V_{\mathrm{ref}}$ are directly applied by the VSC without modification. 
 %
The constrained GFM control restricts the unconstrained candidate voltage to the admissible set ${\mathcal{C}}^{\text{pv}} \coloneqq {\mathcal{C}}^{\max}_{\mathrm{i}} \cap \mathcal{C}_{\mathrm{mod}} \cap \mathcal{C}_{\mathrm{dc}}^{\max} \cap \mathcal{C}_{\mathrm{dc}}^{\min}$. If $v_{\mathrm{i,dq}}^{\mathrm{ref}}=(V_{\mathrm{ref}},0)\in{\mathcal{C}}^{\text{pv}}$, then ${\theta}^{\text{dp}}$ and $V_{\mathrm{ref}}$ obtained from the unconstrained control are applied by the VSC without modification. Otherwise, the input is considered infeasible, as it violates one or more safety constraints. In this case, an optimization problem is solved to find a feasible input that minimizes the deviation from the (unconstrained) reference input ${\theta}^{\text{dp}}$ and $V_{\mathrm{ref}}$. Consequently, the closest admissible control input that satisfies all CBF constraints ${\mathcal{C}}^{\text{pv}}$ is applied by the VSC and used to update the controller states ${\theta}^{\text{dp}}$ and $V_{\mathrm{ref}}$.

Given unconstrained GFM voltage references ${\theta}^{\text{dp}}$ and $V_{\mathrm{ref}}$  in polar coordinates and constraint set  ${\mathcal{C}}^{\text{pv}}$, one can write the CBF-based constrained GFM control in Polar coordinates as
\begin{align}\label{eq.CBFpolar}
(V^{\text{cbf}},{\theta}^{\text{cbf}}) = &\mathop{\mathrm{argmin}}_{\hat{V},\hat{\theta}} \;  (\hat{V}-V_{\mathrm{ref}})^2  + \gamma_{\theta}V_{\mathrm{ref}}^{2-\rho}(\hat{\theta}-{\theta}^{\text{dp}})^2 \\
\textrm{s.t. \quad} 
& \mathbf{R}(\hat{\theta}-{\theta}^{\text{dp}}) \left[ \hat{V}, 0 \right]^{\top} \in \mathcal{C}^{\mathrm{pv}},\nonumber
\end{align}
%
% \begin{align}\label{eq.CBFpolar_new}
% (V^{\text{cbf}},{\theta}^{\text{cbf}}) = &\mathop{\mathrm{argmin}}_{\hat{V},\hat{\theta}} \;  V_{\mathrm{ref}}^{\rho}(\hat{V}-V_{\mathrm{ref}})^2  + 2 \gamma_{\theta}(1-\cos(\hat{\theta}-{\theta}^{\text{dp}})) \\
% \textrm{s.t. \quad} 
% & \mathcal{R}({\theta}^{\text{dp}}-\hat{\theta}) \left[ \hat{V}, 0 \right]^{\top} \in \mathcal{C}^{\mathrm{pv}},\nonumber
% \end{align}
% %
% \begin{align}\label{eq.CBFpolar_new_no_singularity}
% (V^{\text{cbf}},{\theta}^{\text{cbf}}) = &\mathop{\mathrm{argmin}}_{\hat{V},\hat{\theta}} \;  V_{\mathrm{ref}}^{\rho}(\hat{V}-V_{\mathrm{ref}})^2  + 2 \gamma_{\theta}V_{\mathrm{ref}}^{2}(1-\cos(\hat{\theta}-{\theta}^{\text{dp}})) \\
% \textrm{s.t. \quad} 
% & \mathcal{R}({\theta}^{\text{dp}}-\hat{\theta}) \left[ \hat{V}, 0 \right]^{\top} \in \mathcal{C}^{\mathrm{pv}},\nonumber
% \end{align}
%
where $\gamma_{\theta} \in \mathbb{R}_{>0}$ is a weight that determines the trade-off between modifying the GFM voltage magnitude and phase angle. Moreover, $\rho \in \mathbb{R}_{>0}$ adjusts the cost of phase-angle modifications during low-voltage contingencies. Notably, $\rho = 0$ penalizes arc length instead of phase angle modifications, $\rho = 2$ penalizes phase angle modifications independently of the voltage magnitude, and $\rho > 2$ results in a penalty on phase angle modifications that increases as $V_{\mathrm{ref}} \to 0$. Angle differences in \eqref{eq.CBFpolar} are taken modulo $2\pi$ and represented in $[-\pi,\pi)$. It is worth noting that the equations in the $\text{dq}$ reference frame used to formulate the constrained set ${\mathcal{C}}^{\text{pv}}$ are obtained by applying a nonlinear coordinate transformation parametrized by ${\theta}^{\text{dp}}$. Thus, \eqref{eq.CBFpolar} is a nonlinear, nonconvex problem that is challenging to solve at the sampling times required for GFM control.

Therefore, to make the constrained GFM controller computation tractable in real time, we approximate \eqref{eq.CBFpolar} as a convex problem. Considering the relationship ${\theta}^{\text{cbf}}-{\theta}^{\text{dp}} = \angle v_{\mathrm{i,dq}}$ and $V^{\text{cbf}} = \norm{v_{\mathrm{i,dq}}}$, we obtain $(\hat{V}-V_{\mathrm{ref}})^2 = (\norm{v_{\mathrm{i,dq}}}- V_{\mathrm{ref}})^2$ and $\gamma_{\theta}(\hat{\theta}-{\theta}^{\text{dp}})^2 = \gamma_{\theta}(\angle v_{\mathrm{i,dq}})^2$. Then, a second-order approximation of these terms around the operating point $v_{\mathrm{i,dq}}^{\text{ref}} = ( V_{\mathrm{ref}}, 0)$ allows us to approximate \eqref{eq.CBFpolar} as

\begin{align}\label{eq.CBFcartes}
(v_{\mathrm{id}}^{\text{cbf}},v_{\mathrm{iq}}^{\text{cbf}}) = &\mathop{\mathrm{argmin}}_{\hat{v}_{\mathrm{id}}^{\mathrm{cbf}},\hat{v}_{\mathrm{iq}}^{\mathrm{cbf}}} \;  (\hat{v}_{\mathrm{id}}^{\text{cbf}}-V_{\mathrm{ref}})^2  + \frac{\gamma_{\theta}}{{V_{\mathrm{ref}}}^\rho}(\hat{v}_{\mathrm{iq}}^{\text{cbf}})^2 \\
\textrm{s.t. \quad} 
& (\hat{v}_{\mathrm{id}}^{\text{cbf}},\hat{v}_{\mathrm{iq}}^{\text{cbf}}) \in \mathcal{C}^{\mathrm{pv}},\nonumber
\end{align}

Note that the constraint sets $ {\mathcal{C}}^{\max}_{\mathrm{i}}$, $\mathcal{C}_{\mathrm{dc}}^{\max}$, and $\mathcal{C}_{\mathrm{dc}}^{\min}$ are represented by two-dimensional half-spaces, whereas the modulation constraint $\mathcal{C}_{\mathrm{mod}}$ is represented by a circle centered at the origin. Consequently, the resulting feasible set ${\mathcal{C}}^{\text{pv}}$ is convex. Moreover, the voltage phase angle and magnitude in polar coordinates can be recovered as $\theta^{\mathrm{cbf}}=\theta^{\mathrm{dp}}+\angle v_{\mathrm{i,dq}}^{\mathrm{cbf}}$ and $V^{\mathrm{cbf}}=\|v_{\mathrm{i,dq}}^{\mathrm{cbf}}\|$, respectively.

Discretizing the GFM controller using the forward-Euler method with a sampling period of $\tau_{\text{c}}$ and discrete-time index $k \in \mathbb{Z}_{\ge 0}$ results in
%
%\begin{subequations}\label{eq.dualPort2} 
%	\begin{align}
	% \ddt \theta^{\text{dp}} &=  \omega^{\star} + \frac{{k_\mathrm{p}} \tau_{\mathrm{p}}}{{\tau}_{\mathrm{pv}}}v_{{\delta},\mathrm{dc}} + {k_\mathrm{p}}\left(  k_{\mathrm{pv}} - \frac{\tau_{\mathrm{p}}}{{\tau}_{\mathrm{pv}}} \right)\lambda,\\
	% {\tau}_{\mathrm{pv}} \ddt \lambda &= - \lambda + v_{{\delta},\mathrm{dc}},
	%\end{align}
	%\end{subequations}
%	\begin{subequations}\label{eq.dualPort2} 
		\begin{align}\label{eq.dualPort2} 
			\theta^{\text{dp}}\left(k\!+\!1\right) \!=\!& \; \theta^{\text{cbf}}(k) + \tau_{\text{c}} \omega^{\star} + \tfrac{\tau_{\text{c}} k_{\text{p}} \tau_{\text{p}}}{\tau_{\text{pv}}}v_{{\delta},\mathrm{dc}}(k)\\ \nonumber
			&+ \tau_{\text{c}} k_{\text{p}}\left(  k_{\mathrm{pv}} - \tfrac{\tau_{\mathrm{p}}}{{\tau}_{\mathrm{pv}}} \right)\lambda(k),\\ \nonumber
			\lambda \left(k\!+\!1\right) \!=\!& \left(1- \tfrac{\tau_{\text{c}}}{\tau_{\text{pv}}} \right) \lambda (k) + \tfrac{\tau_{\text{c}}}{\tau_{\text{pv}}}v_{{\delta},\mathrm{dc}}(k),\\
					V_{\mathrm{ref}} (k\!+\!1) \!=\!& \left(1\!-\! \tfrac{\tau_{\text{c}}}{\tau_{\text{v}}} \right) V^{\mathrm{cbf}}(k) + \tfrac{\tau_{\text{c}}}{\tau_{\text{v}}}\left(V^{\star} \!+\! m_{\mathrm{q}} ( Q^{\star} \!-\! \tilde{Q}(k))\right)\!,\nonumber
		\end{align}
%	\end{subequations}
	where $\lambda \in\mathbb{R}$ is the state of a lowpass filter and 
	${\theta}^{\text{dp}} \in [-\pi, \pi)$ is the GFM voltage phase angle.

\begin{figure*} 
	\centerline{\includegraphics[scale=0.22,trim={14cm 2cm 10cm 6cm}]{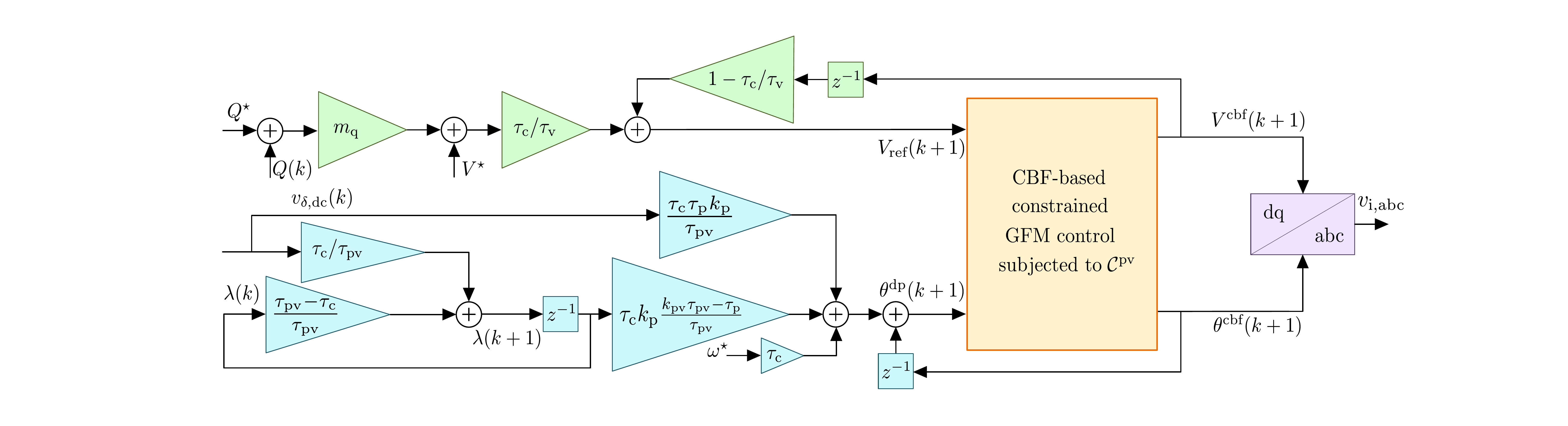}}
	\caption{Discrete-time implementation of the proposed constrained GFM control strategy with sampling period $\tau_{\mathrm{c}}$. The reference signals are generated by discrete-time control \eqref{eq.dualPort2} and projected onto $\mathcal{C}^{\text{pv}}$ according to \eqref{eq.CBFcartes}.}
	\label{fig.CBFblocks}
\end{figure*}

The discrete-time implementation of the resulting constrained GFM control is shown in Fig.~\ref{fig.CBFblocks}. At every controller time-step $k \in \mathbb{N}_{>0}$, one time step of the unconstrained GFM control starting from  $V^{\text{cbf}}(k)$ and $\theta^{\text{cbf}}(k)$ is used to compute the unconstrained candidate voltage $(V_{\mathrm{ref}}, 0)$ and ${\theta}^{\text{dp}}$. Moreover, ${\theta}^{\text{dp}}$ is used to construct the constraint set ${\mathcal{C}}^{\text{pv}}$ in the dq-frame of the unconstrained GFM reference. If the candidate voltage $v_{\mathrm{i,dq}}^{\text{ref}}$ is within the safe set ${\mathcal{C}}^{\text{pv}}$ then the optimal solution of \eqref{eq.CBFcartes} is $(V_{\mathrm{ref}}, 0)$ and the control is identical to dual-port GFM control, i.e., the angle and voltage magnitude obtained from \eqref{eq.dualPort2} are used without any modifications. In contrast, if $(V_{\mathrm{ref}}, 0)$ is outside the safe set ${\mathcal{C}}^{\text{pv}}$, then the closest admissible voltage magnitude $V^{\text{cbf}}$ and $\theta^{\text{cbf}}$ are computed using \eqref{eq.CBFcartes}.

\subsection{Real-time optimization}
To efficiently solve \eqref{eq.CBFcartes}, we use the alternating direction method of multipliers (ADMM)~\cite[Sec.~3.1]{neal2011distributed} inspired by~\cite[Sec.~IV-C]{GrossConstrained}. At each controller time step $k \in \mathbb{N}$, $n_{\mathrm{itr}}\in \mathbb{N}_{>0}$ iterations of the ADMM algorithm are performed during which the unconstrained voltage references and constraint sets are held constant.

Let $W\coloneqq 2\operatorname{diag}\!\left(1,\gamma_{\theta}/(V_{\mathrm{ref}})^{\rho}\right)$, where $V_{\mathrm{ref}}>0$. For brevity of the notation, define $\mathcal{C}_1\coloneqq\mathcal{C}_{\mathrm{i}}^{\max}$, $\mathcal{C}_2\coloneqq\mathcal{C}_{\mathrm{dc}}^{\max}$, $\mathcal{C}_3\coloneqq\mathcal{C}_{\mathrm{dc}}^{\min}$, and $\mathcal{C}_4\coloneqq\mathcal{C}_{\mathrm{mod}}$. Introducing the splitting variables $z_j\in\mathbb{R}^2$, $j\in\{1,2,3,4\}$, \eqref{eq.CBFcartes} can be equivalently written as
\begin{subequations}\label{eq.admmConsensus}
\begin{align}
\min_{\hat{v}_{\mathrm{i,dq}}^{\mathrm{cbf}},z_1,\ldots,z_4}\quad
&\tfrac{1}{2}(\hat{v}_{\mathrm{i,dq}}^{\mathrm{cbf}}-v_{\mathrm{i,dq}}^{\mathrm{ref}})^{\top}W(\hat{v}_{\mathrm{i,dq}}^{\mathrm{cbf}}-v_{\mathrm{i,dq}}^{\mathrm{ref}})\\
\text{s.t.}\quad &\hat{v}_{\mathrm{i,dq}}^{\mathrm{cbf}}=z_j,\quad z_j\in\mathcal{C}_j,\quad j\in\{1,2,3,4\}.
\end{align}
\end{subequations}

Next, we introduce the penalty $\sigma>0$, $W_{\sigma}\coloneqq W+4\sigma I_2$, and the scaled dual multipliers $y_j\in\mathbb{R}^2$ associated with the constraint $\hat{v}_{\mathrm{i,dq}}^{\mathrm{cbf}}=z_j$. The ADMM algorithm is given by
\begin{subequations}\label{eq.ADMM}
\begin{align}
\hat{v}_{\mathrm{i,dq}}^{\mathrm{cbf},n+1}&=W_{\sigma}^{-1}\!\bigg(Wv_{\mathrm{i,dq}}^{\mathrm{ref}}
+\sigma\sum\nolimits_{j=1}^{4}(z_j^{n}-y_j^{n})\bigg),\label{eq.ADMMx}\\
z_j^{n+1}&=\Pi_{\mathcal{C}_j}(\hat{v}_{\mathrm{i,dq}}^{\mathrm{cbf},n+1}+y_j^{n}),\label{eq.ADMMz}\\
y_j^{n+1}&=y_j^{n}+\hat{v}_{\mathrm{i,dq}}^{\mathrm{cbf},n+1}-z_j^{n+1}.\label{eq.ADMMy}
\end{align}
\end{subequations}

Here, the updates \eqref{eq.ADMMz} and \eqref{eq.ADMMy} for every constraint $j\in\{1,2,3,4\}$ can be performed in parallel and leverage the Euclidean projection $\Pi_{\mathcal{C}_j}(\xi)$ of $\xi \in \mathbb{R}^2$ onto $\mathcal{C}_j$. To this end, let $a_1=a_3=i_{\mathrm{f,dq}}$, $a_2=-i_{\mathrm{f,dq}}$, $b_1=\alpha_{\mathrm{i}}^{\max}h_{\mathrm{i}}^{\max}+P_{\mathrm{out}}+P_{\mathrm{loss}}$, $b_2=\alpha_{\mathrm{dc}}^{\max}h_{\mathrm{dc}}^{\max}-P_{\mathrm{dc}}$, and $b_3=\alpha_{\mathrm{dc}}^{\min}h_{\mathrm{dc}}^{\min}+P_{\mathrm{dc}}$. Then, for $j\in\{1,2,3\}$, it holds that 
\begin{align}\label{eq.admmHalfspaceProjection}
\Pi_{\mathcal{C}_j}(\xi)
=\xi-\frac{\max\{a_j^{\top}\xi-b_j,0\}}{\norm{a_j}^2}a_j
\end{align}
if $a_j\ne0_2$, $\Pi_{\mathcal{C}_j}(\xi)=\xi$ if $a_j=0_2$ and $b_j\ge0$, and the set is empty otherwise (i.e., a constraint violation already occurred). Moreover, for the modulation limit $j=4$, $\Pi_{\mathcal{C}_{j}}(\xi)=\xi$ if $\norm{\xi}\le V^{\max}_{\mathrm{mod}}$ and $\Pi_{\mathcal{C}_{\mathrm{mod}}}(\xi)=V^{\max}_{\mathrm{mod}}\xi/\norm{\xi}$ otherwise. Crucially, the algorithm \eqref{eq.ADMM} alternates between the averaging step \eqref{eq.ADMMx} and the projections \eqref{eq.ADMMz} onto the constraint sets and is computationally lightweight because closed-form solutions are available for the Euclidean projections onto the constraint sets. At every controller time step $k \in \mathbb{N}_{>0}$ the variables are initialized using the unconstrained GFM voltage reference, i.e., $z_j^0=v_{\mathrm{i,dq}}^{\mathrm{ref}}=(V_{\mathrm{ref}},0)^{\top}$ and $y_j^0=0_2$ for all $j\in\{1,2,3,4\}$, and a fixed number of iterations $n_{\mathrm{itr}}$ of \eqref{eq.ADMM} are performed.

\section{Case Study: Multi-Source System}
\begin{figure}[b!!!]  
  \centerline{\includegraphics[scale=0.24,trim={8cm 2.2cm 10cm 3cm}]{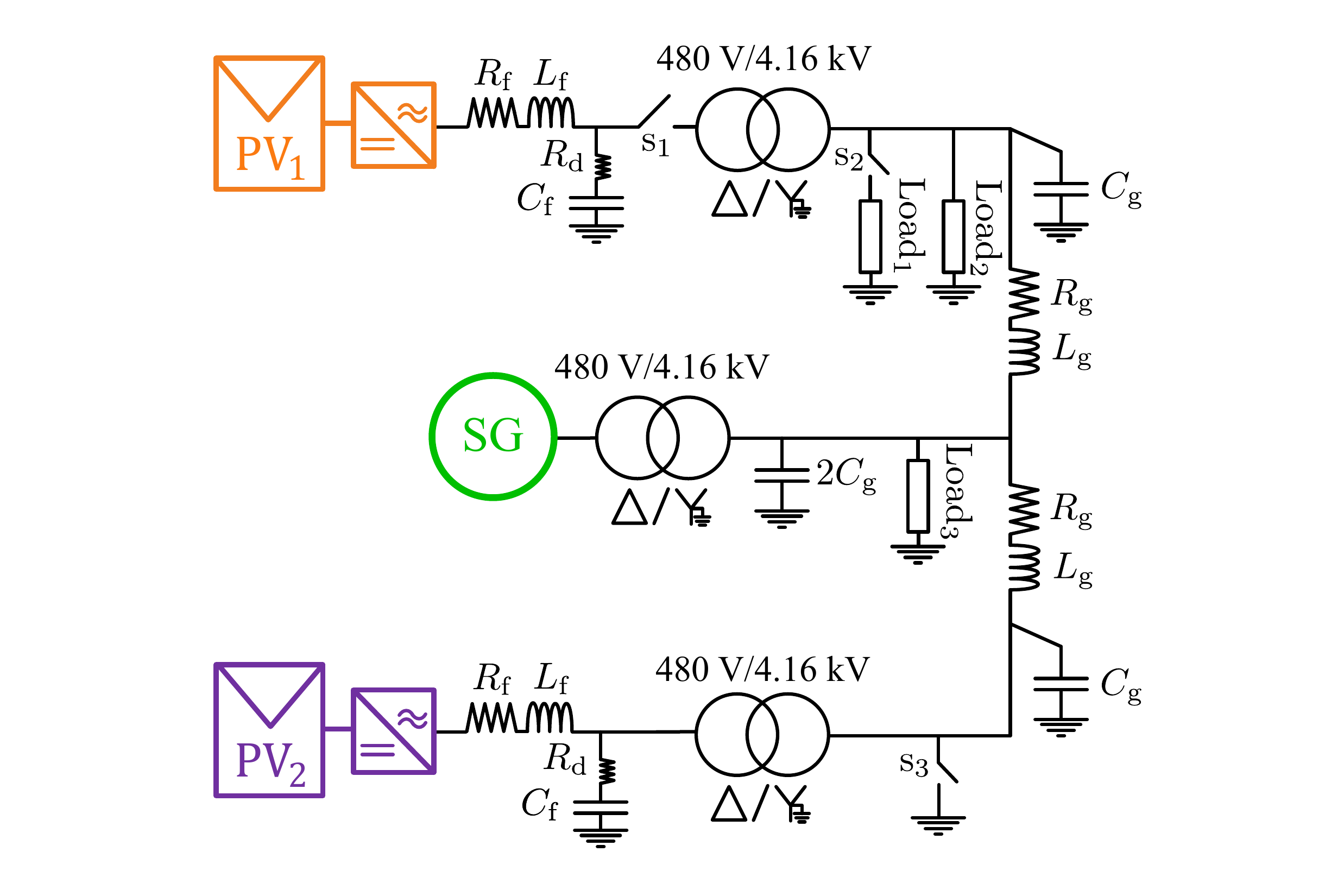}}
  \caption{Test system with two GFM single-stage PV units and one SG.}
  \label{fig.grid}
  \end{figure}

The multi-unit system shown in Fig.~\ref{fig.grid} is used to evaluate the performance of the constrained GFM control.
\subsection{System Model}
The multi-source system includes two single-stage GFM PV systems and one synchronous generator. To model a $1~\mathrm{MVA}$ single-stage PV unit with a central inverter, we use an averaged model of a two-level VSC and $104$ parallel strings with $32$ series-connected AU Optronics $\text{PM}060\text{MBR}$ modules per string. The maximum and minimum voltages are functions of the irradiance and temperature. For this study, the base dc voltage for PVs is considered to be equal to $v_{\mathrm{dc}}^{\text{mpp}}$, which is $1060$~V for an irradiance of $1000$~$\mathrm{W/m^2}$. The maximum dc-bus voltage is $v_{\mathrm{dc}}^{\text{oc}} = 1310$~V.  The maximum current magnitude for both PVs is selected to be $1.2$~p.u. The voltage droop constant of the PV units is $m_{\mathrm{q}} = 0.03$~p.u. The SG is simulated using a sixth-order nonlinear model with a first-order turbine/droop governor model with time constant $\tau_{\mathrm{g}}$. The SG also includes a multiband power system stabilizer and an IEEE AC1A excitation-system model with an automatic voltage regulator. The relevant system, SG, and VSC parameters are summarized in Table~\ref{tab.Multiparameters}. The system is modeled in MATLAB/Simulink using the Specialized Power Systems library, with a simulation time step of $10$~$\mu$s. The remainder of this section uses various events and contingencies to test the response of the GFM PV systems under constraints.

 \begin{table}
\caption{Multi-Source System Parameters}
\label{tab.Multiparameters}
\centering
\small
\setlength{\tabcolsep}{2pt}
\begin{tabular*}{\columnwidth}{@{\extracolsep{\fill}}clclcl@{}}
\toprule
\multicolumn{2}{c}{Grid base values} &
\multicolumn{2}{c}{VSC base values} &
\multicolumn{2}{c}{SG}\\
\midrule
$S_{\mathrm{b}}$ & $1.5$~MW &
$S_{\mathrm{b}}$ & $1$~MW &
$\tau_{\mathrm{g}}$ & $1$~s\\
$V_{\mathrm{b}}$ & $4.16$~kV &
$V_{\mathrm{b}}$ & $480$~V &
$H$ & $0.8$~s\\
$f_{\mathrm{b}}$ & $60$~Hz &
$f_{\mathrm{b}}$ & $60$~Hz &
$m_{\mathrm{p}}$ & $0.03$~p.u.\\
& &
$v_{\mathrm{dc,b}}$ & $v_{\mathrm{dc}}^{\mathrm{mpp}}$ &
&\\
\midrule
\multicolumn{2}{c}{VSC circuit} &
\multicolumn{2}{c}{Network circuit} &
\multicolumn{2}{c}{Control parameters}\\
\midrule
$L_{\mathrm{f}}$ & $0.1$~p.u. &
$L_{\mathrm{g}}$ & $0.055$~p.u. &
$\tau_{\mathrm{v}}$ & $8$~ms\\
$R_{\mathrm{f}}$ & $0.01$~p.u. &
$R_{\mathrm{g}}$ & $0.018$~p.u. &
$\tau_{\mathrm{p}}$ & $0.56$~p.u.\\
$R_{\mathrm{d}}$ & $0.2$~p.u. &
$C_{\mathrm{g}}$ & $4.35{\times}10^{-5}$~p.u. &
$k_{\mathrm{p}}$ & $0.05$~p.u.\\
$C_{\mathrm{f}}$ & $0.05$~p.u. &
$L_{\mathrm{T}}$ & $0.03$~p.u. &
$\tau_{\mathrm{pv}}$ & $0.01$~s\\
$C_{\mathrm{dc}}$ & $0.4$~F &
$R_{\mathrm{T}}$ & $0.002$~p.u. &
$\gamma_{\theta}$ & $0.13$~p.u.\\
$\tau_{\mathrm{e}}$ & $0.1$~s &
& &
$\rho$ & $5$\\
\midrule
\multicolumn{2}{c}{Load} &
\multicolumn{2}{c}{CBF} &
\multicolumn{2}{c}{ADMM}\\
\midrule
$\mathrm{Load}_{1}$ & $0.63$~p.u. &
$\alpha_{\mathrm{i}}^{\max}$ & $32$~p.u. &
$n_{\mathrm{itr}}$ & $5$\\
$\mathrm{Load}_{2}$ & $0.27$~p.u. &
$\alpha_{\mathrm{dc}}^{\max}$ & $20$~p.u. &
$\sigma$ & $3$\\
$\mathrm{Load}_{3}$ & $0.55$~p.u. &
$\alpha_{\mathrm{dc}}^{\min}$ & $20$~p.u. &
&\\
\bottomrule
\end{tabular*}
\end{table}

 \begin{figure}[b!!]
  \centerline{\includegraphics[scale=0.39,trim={52cm 0.6cm 53cm 0.5cm}]{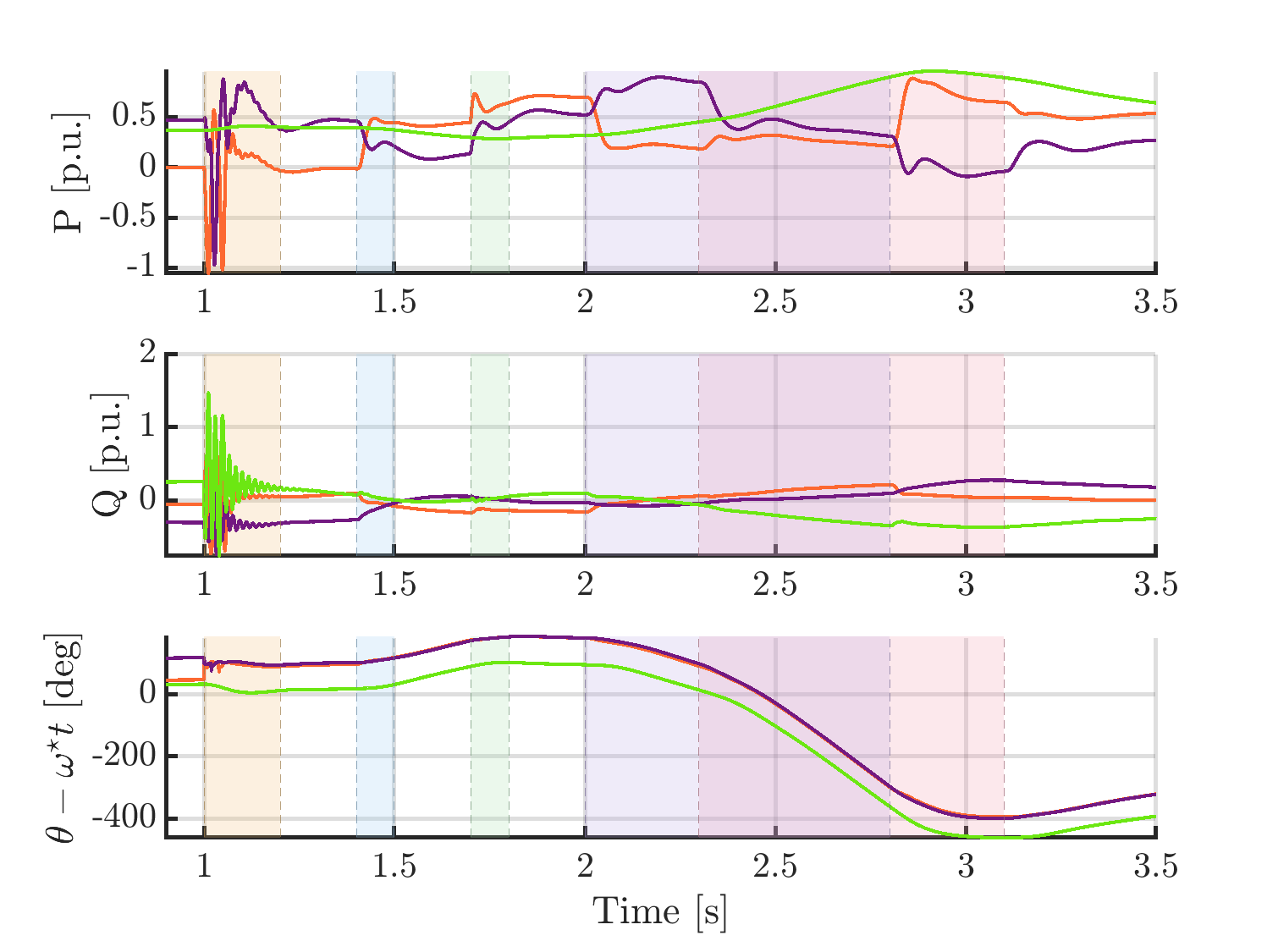}}
  \caption{Active powers, reactive powers, and phase angles of $\text{PV}_{\mathrm{1}}$ (\textcolor[rgb]{0.9882, 0.4078, 0.1922}{\rule[0.5ex]{1.2em}{0.6pt}}),  $\text{PV}_{\mathrm{2}}$ (\textcolor[rgb]{0.4510, 0.0980, 0.5059}{\rule[0.5ex]{1.2em}{0.6pt}}), and the SG (\textcolor[rgb]{0.4235, 0.9059, 0.0706}{\rule[0.5ex]{1.2em}{0.6pt}}) for the synchronization event {\setlength{\fboxsep}{1pt}%
  	(\colorbox[rgb]{0.9890,0.9450,0.8790}{\rule{0pt}{1.2ex}\hspace{0.9em}})},  dc voltage setpoint change {\setlength{\fboxsep}{1pt}%
  	(\colorbox[rgb]{0.9120,0.9560,0.9890}{\rule{0pt}{1.2ex}\hspace{0.9em}})}
  	,  load step {\setlength{\fboxsep}{1pt}%
  		(\colorbox[rgb]{0.9230,0.9736,0.9296}{\rule{0pt}{1.2ex}\hspace{0.9em}})}
  	, and $\text{PV}_{\mathrm{1}}$ {\setlength{\fboxsep}{1pt}%
  		(\colorbox[rgb]{0.9296,0.9208,0.9780}{\rule{0pt}{1.2ex}\hspace{0.9em}})}
  	 and $\text{PV}_{\mathrm{2}}$ irradiance drop {\setlength{\fboxsep}{1pt}%
  		(\colorbox[rgb]{0.9890,0.9164,0.9010}{\rule{0pt}{1.2ex}\hspace{0.9em}})}. }
  \label{fig.PQOtherScen}
  \end{figure}

\subsection{Resynchronization, Load Step, and Irradiance Change}
The system initially operates with switches $\mathrm{s}_1$, $\mathrm{s}_2$, and $\mathrm{s}_3$ open and $\text{PV}_{\mathrm{1}}$ disconnected from the grid. The dc-link voltage operating points are initially selected as $v_{\mathrm{dc,1}}^{\star} = v_{\mathrm{dc,1}}^{\mathrm{oc}} = 1.23$~p.u. (i.e., zero power generation) and $v_{\mathrm{dc,2}}^{\star} = 1.17$~p.u. At $t=1$~s, the switch $\mathrm{s}_1$ is closed to connect $\text{PV}_{\mathrm{1}}$ to the system. It can be observed from Fig.~\ref{fig.PQOtherScen} that $\text{PV}_{\mathrm{1}}$ initially has a voltage phase-angle difference of approximately $50^{\circ}$ with respect to $\text{PV}_{\mathrm{2}}$ and the SG, representing a severe lack of synchronization.

\begin{figure}
  \centerline{\includegraphics[scale=0.39,trim={52cm 0.6cm 53cm 0.5cm}]{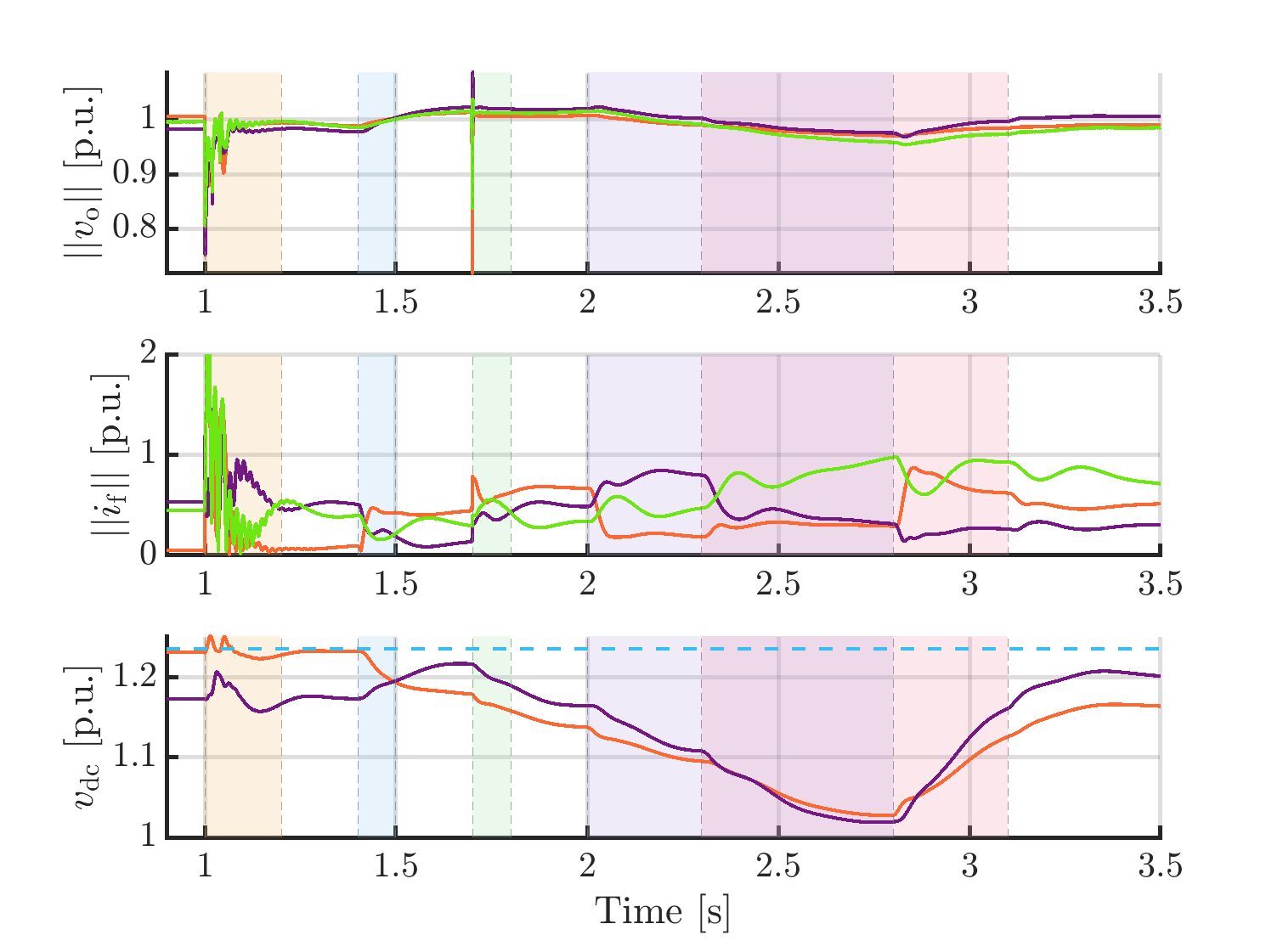}}
  \caption{
  Terminal voltage magnitude, current magnitudes of $\text{PV}_{\mathrm{1}}$ (\textcolor[rgb]{0.9882, 0.4078, 0.1922}{\rule[0.5ex]{1.2em}{0.6pt}}),  $\text{PV}_{\mathrm{2}}$ (\textcolor[rgb]{0.4510, 0.0980, 0.5059}{\rule[0.5ex]{1.2em}{0.6pt}}), and the SG (\textcolor[rgb]{0.4235, 0.9059, 0.0706}{\rule[0.5ex]{1.2em}{0.6pt}}), dc-link voltages and maximum dc voltage (\textcolor[rgb]{0.2235, 0.7412, 0.9294}{\rule[0.5ex]{0.32em}{0.6pt}\hspace{0.12em}\rule[0.5ex]{0.32em}{0.6pt}\hspace{0.12em}\rule[0.5ex]{0.32em}{0.6pt}}) for the synchronization event {\setlength{\fboxsep}{1pt}%
  	(\colorbox[rgb]{0.9890,0.9450,0.8790}{\rule{0pt}{1.2ex}\hspace{0.9em}})},  dc voltage setpoint change {\setlength{\fboxsep}{1pt}%
  	(\colorbox[rgb]{0.9120,0.9560,0.9890}{\rule{0pt}{1.2ex}\hspace{0.9em}})}
  	,  load step {\setlength{\fboxsep}{1pt}%
  		(\colorbox[rgb]{0.9230,0.9736,0.9296}{\rule{0pt}{1.2ex}\hspace{0.9em}})}
  	, and $\text{PV}_{\mathrm{1}}$ {\setlength{\fboxsep}{1pt}%
  		(\colorbox[rgb]{0.9296,0.9208,0.9780}{\rule{0pt}{1.2ex}\hspace{0.9em}})}
  	 and $\text{PV}_{\mathrm{2}}$ irradiance drop {\setlength{\fboxsep}{1pt}%
  		(\colorbox[rgb]{0.9890,0.9164,0.9010}{\rule{0pt}{1.2ex}\hspace{0.9em}})}. }
  \label{fig.viOtherScen}
  \end{figure}

\begin{figure}[b]
  \centerline{\includegraphics[scale=0.39,trim={52cm 0.6cm 53cm 0.5cm}]{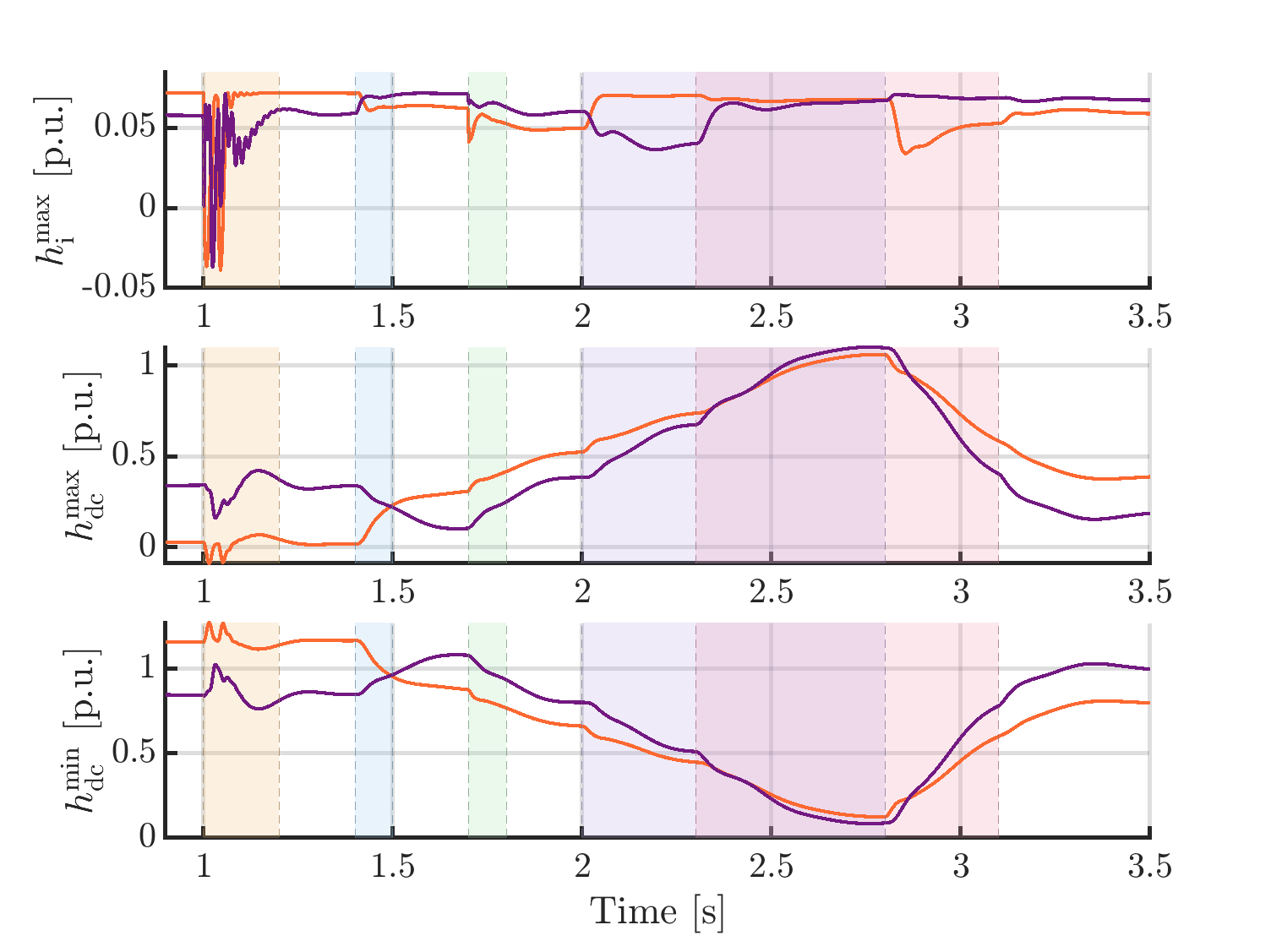}}
  \caption{CBFs encoding current limits and dc-link voltage limits for $\text{PV}_{\mathrm{1}}$ (\textcolor[rgb]{0.9882, 0.4078, 0.1922}{\rule[0.5ex]{1.2em}{0.6pt}}) and  $\text{PV}_{\mathrm{2}}$ (\textcolor[rgb]{0.4510, 0.0980, 0.5059}{\rule[0.5ex]{1.2em}{0.6pt}}) for the synchronization event {\setlength{\fboxsep}{1pt}%
  	(\colorbox[rgb]{0.9890,0.9450,0.8790}{\rule{0pt}{1.2ex}\hspace{0.9em}})},  dc voltage setpoint change {\setlength{\fboxsep}{1pt}%
  	(\colorbox[rgb]{0.9120,0.9560,0.9890}{\rule{0pt}{1.2ex}\hspace{0.9em}})}
  	,  load step {\setlength{\fboxsep}{1pt}%
  		(\colorbox[rgb]{0.9230,0.9736,0.9296}{\rule{0pt}{1.2ex}\hspace{0.9em}})}
  	, and $\text{PV}_{\mathrm{1}}$ {\setlength{\fboxsep}{1pt}%
  		(\colorbox[rgb]{0.9296,0.9208,0.9780}{\rule{0pt}{1.2ex}\hspace{0.9em}})}
  	 and $\text{PV}_{\mathrm{2}}$ irradiance drop {\setlength{\fboxsep}{1pt}%
  		(\colorbox[rgb]{0.9890,0.9164,0.9010}{\rule{0pt}{1.2ex}\hspace{0.9em}})}. }
  \label{fig.HOtherScen}
  \end{figure}

The synchronization of $\text{PV}_{\mathrm{1}}$ causes fast transient oscillations in the system, as observed in Figs.~\ref{fig.PQOtherScen},~\ref{fig.viOtherScen}, and~\ref{fig.HOtherScen}. Nevertheless, the system withstands this severe contingency. $\text{PV}_{\mathrm{1}}$ successfully resynchronizes, its current exhibits only small and brief transient excursions beyond the long-term current limit, and the system returns to its pre-event operating point.

Subsequently, at $t=1.4$~s, the dc voltage setpoint of $\text{PV}_{\mathrm{1}}$ is updated from $v_{\mathrm{dc,1}}^{\star} = v_{\mathrm{dc,1}}^{\mathrm{oc}}$ to $v_{\mathrm{dc,1}}^{\star} = 1.14$~p.u. (i.e., between $v_{\mathrm{dc,1}}^{\mathrm{oc}}$ and $v_{\mathrm{dc,1}}^{\mathrm{mpp}}$) such that $\text{PV}_{\mathrm{1}}$ starts to generate power. Consequently, the power injection of $\text{PV}_{\mathrm{2}}$ decreases and its dc voltage increases to maintain the balance between supply and demand. Note that the SG physics and control result in slow response to the change in dispatch.

Next, at $t=1.7$~s, a large load step is simulated by connecting $\text{Load}_{\mathrm{1}} = 0.63$~p.u. to the grid by closing $\mathrm{s}_2$. As a result, the dc voltages of $\text{PV}_{\mathrm{1}}$ and $\text{PV}_{\mathrm{2}}$ decrease while their power generation immediately increases to supply the increased power demand. Because $\text{PV}_{\mathrm{1}}$ and $\text{PV}_{\mathrm{2}}$ immediately pick up the increased load, the SG does not exhibit a significant response. 

\begin{figure}
  	\centerline{\includegraphics[scale=0.39,trim={52cm 0.6cm 53cm 0.5cm}]{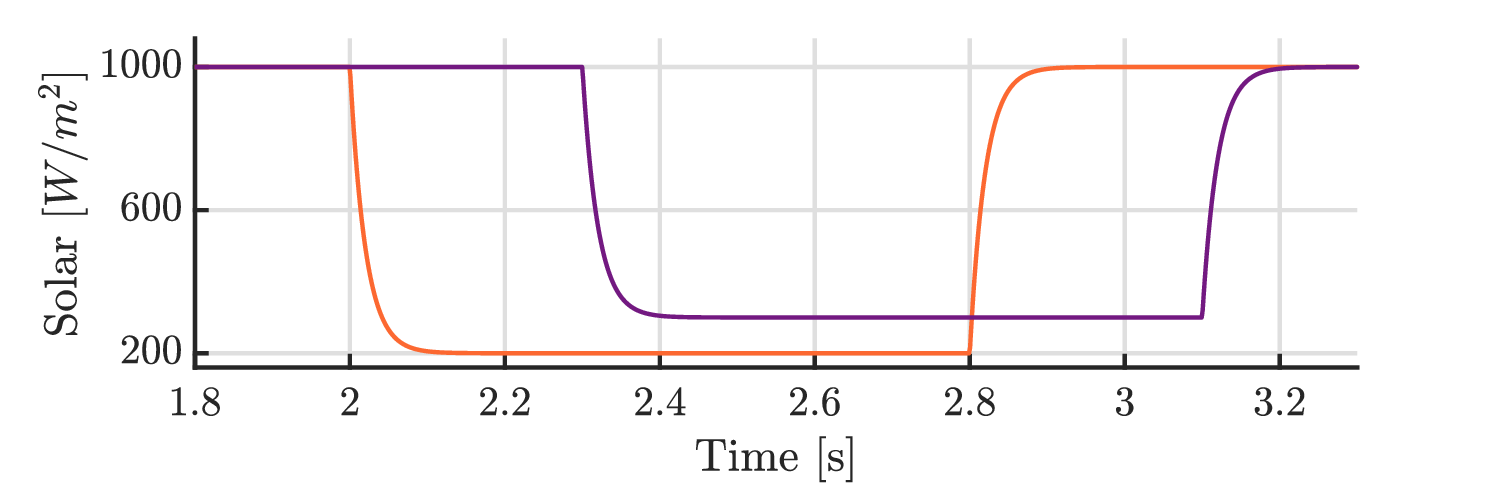}}
  	\caption{Solar irradiance profiles of $\text{PV}_{\mathrm{1}}$ (\textcolor[rgb]{0.9882, 0.4078, 0.1922}{\rule[0.5ex]{1.2em}{0.6pt}}) and  $\text{PV}_{\mathrm{2}}$ (\textcolor[rgb]{0.3490, 0.1569, 0.9294}{\rule[0.5ex]{1.2em}{0.6pt}}).}
  	\label{fig.solar}
  \end{figure}

Finally, we simulate a severe dip in solar irradiance for both PV units to evaluate the response of the proposed control under rapid solar irradiance changes that severely reduce the available PV power generation. Specifically, the solar irradiance of $\text{PV}_{\mathrm{1}}$ drops from $1000$~$\mathrm{W/m^2}$ to $200$~$\mathrm{W/m^2}$ at $t=2$~s, while the solar irradiance of $\text{PV}_{\mathrm{2}}$ drops from $1000$~$\mathrm{W/m^2}$ to $300$~$\mathrm{W/m^2}$ at $t=2.3$~s. The irradiance profiles are depicted in Fig.~\ref{fig.solar}. This rapid drop in irradiance may, in abstraction, model fast-moving clouds passing over the PV arrays. As a result, the total available generation capacity of the system decreases significantly, leading to noticeable drops in the dc voltage and frequency as the GFM PVs respond to the resulting power imbalance. The irradiances subsequently return to their nominal values at $t=2.8$~s and $t=3.1$~s. Subsequently, the power generation and dc-link voltages smoothly return to their pre-contingency values. 

The key observation from these contingencies is that the GFM single-stage PV units and overall system withstand and recover from severe operating disturbances without losing stability. Throughout these contingencies, the ac-current, dc-voltage, and modulation limits remain satisfied. The nonnegative CBF values in Fig.~\ref{fig.HOtherScen} showcase operation of the GFM PV units within their limits.

\subsection{Short-Circuit Fault}
\begin{figure}
  \centerline{\includegraphics[scale=0.39,trim={52cm 0.6cm 53cm 0.5cm}]{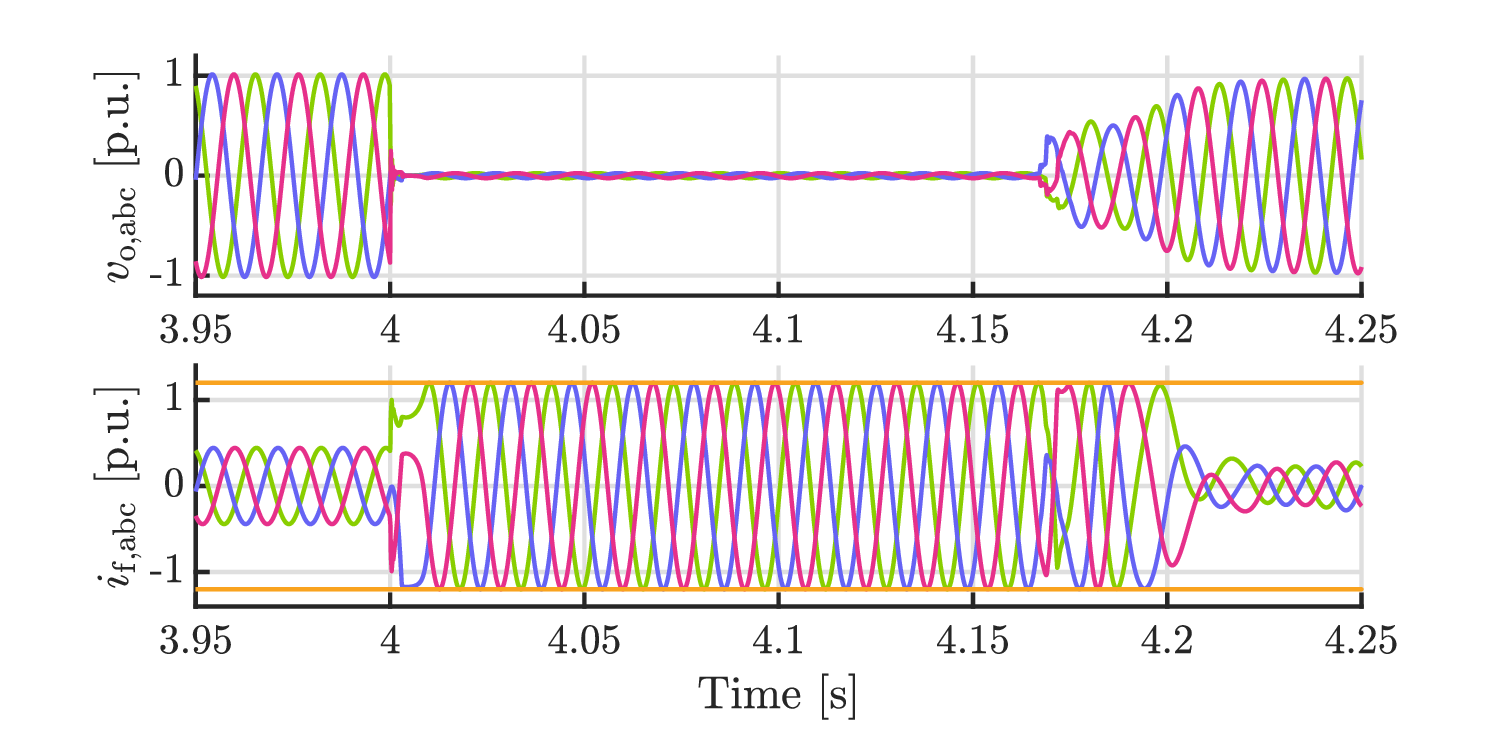}}
  \caption{Phase a (\textcolor[rgb]{0.537, 0.807, 0}{\rule[0.5ex]{1.2em}{0.6pt}}), phase b (\textcolor[rgb]{0.4, 0.388, 0.956}{\rule[0.5ex]{1.2em}{0.6pt}}), and phase c (\textcolor[rgb]{0.90, 0.18, 0.54}{\rule[0.5ex]{1.2em}{0.6pt}}) voltage and current of $\text{PV}_{2}$ during the three-phase short-circuit fault.  \label{fig.abcFault}}
  \end{figure}

\begin{figure}
  \centerline{\includegraphics[scale=0.39,trim={52cm 0.6cm 53cm 0.5cm}]{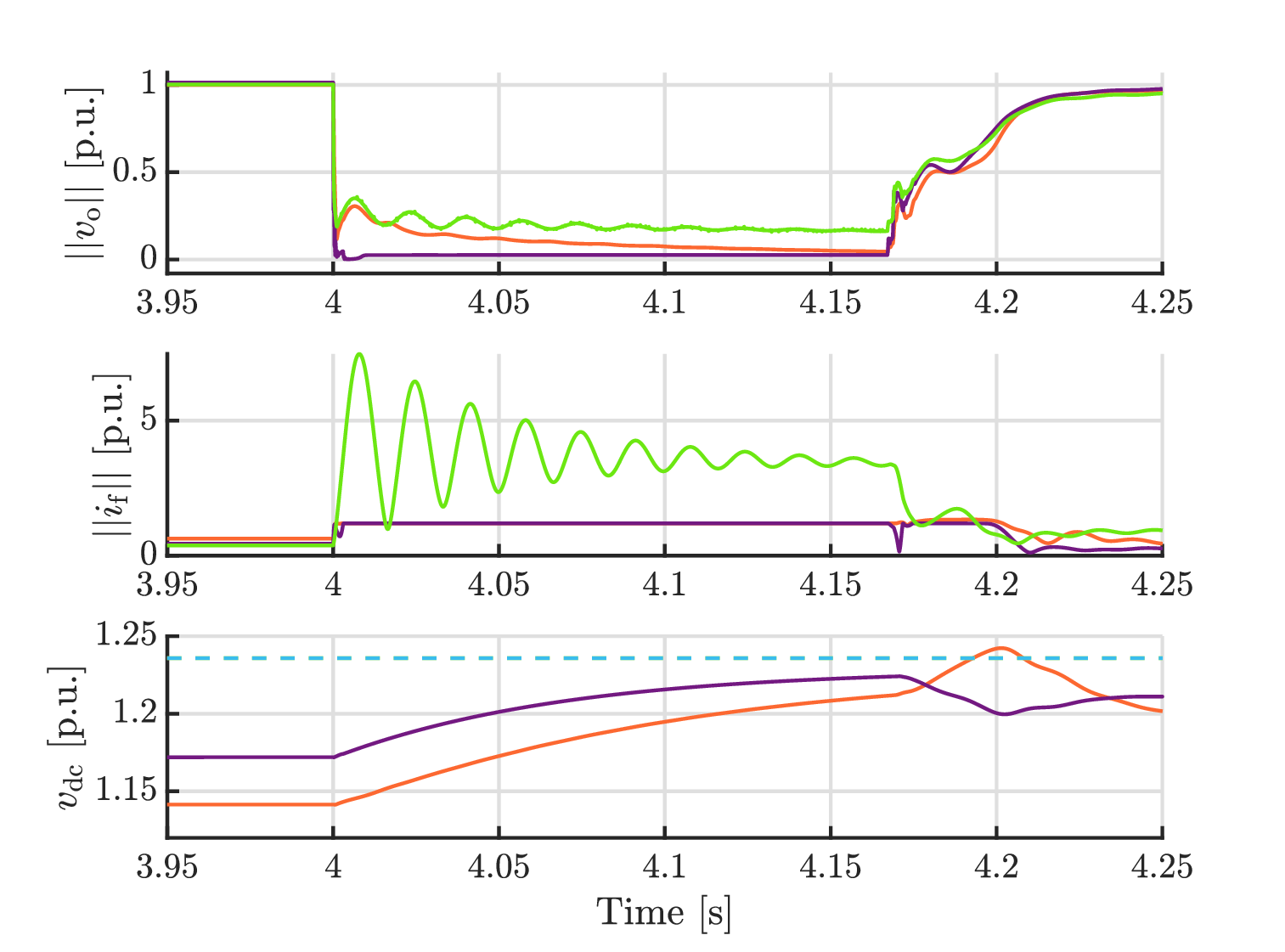}}
  \caption{ AC voltage and current magnitudes of $\text{PV}_{\mathrm{1}}$ (\textcolor[rgb]{0.9882, 0.4078, 0.1922}{\rule[0.5ex]{1.2em}{0.6pt}}),  $\text{PV}_{\mathrm{2}}$ (\textcolor[rgb]{0.4510, 0.0980, 0.5059}{\rule[0.5ex]{1.2em}{0.6pt}}), and the SG (\textcolor[rgb]{0.4235, 0.9059, 0.0706}{\rule[0.5ex]{1.2em}{0.6pt}}), maximum dc voltage (\textcolor[rgb]{0.2235, 0.7412, 0.9294}{\rule[0.5ex]{0.32em}{0.6pt}\hspace{0.12em}\rule[0.5ex]{0.32em}{0.6pt}\hspace{0.12em}\rule[0.5ex]{0.32em}{0.6pt}}), and dc voltages during the short-circuit.  \label{fig.viFault}}
   \end{figure}

\begin{figure}
  \centerline{\includegraphics[scale=0.39,trim={52cm 0.6cm 53cm 0.5cm}]{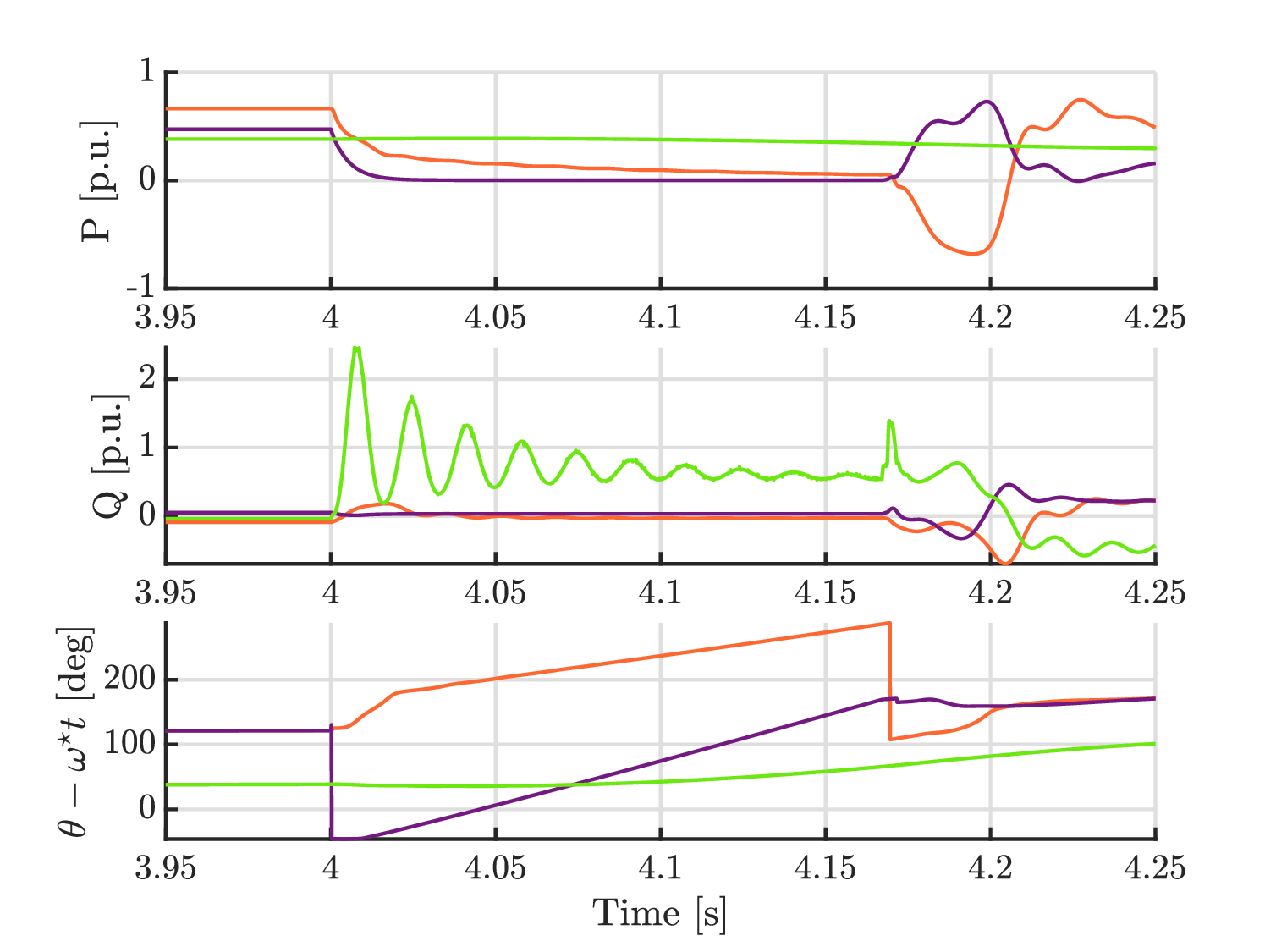}}
  \caption{Active powers, reactive powers, and phase angle difference of $\text{PV}_{\mathrm{1}}$ (\textcolor[rgb]{0.9882, 0.4078, 0.1922}{\rule[0.5ex]{1.2em}{0.6pt}}),  $\text{PV}_{\mathrm{2}}$ (\textcolor[rgb]{0.4510, 0.0980, 0.5059}{\rule[0.5ex]{1.2em}{0.6pt}}), and the SG (\textcolor[rgb]{0.4235, 0.9059, 0.0706}{\rule[0.5ex]{1.2em}{0.6pt}}) during the short-circuit fault.  \label{fig.PQFault}}
  \end{figure}

\begin{figure}
  \centerline{\includegraphics[scale=0.39,trim={52cm 0.6cm 53cm 0.5cm}]{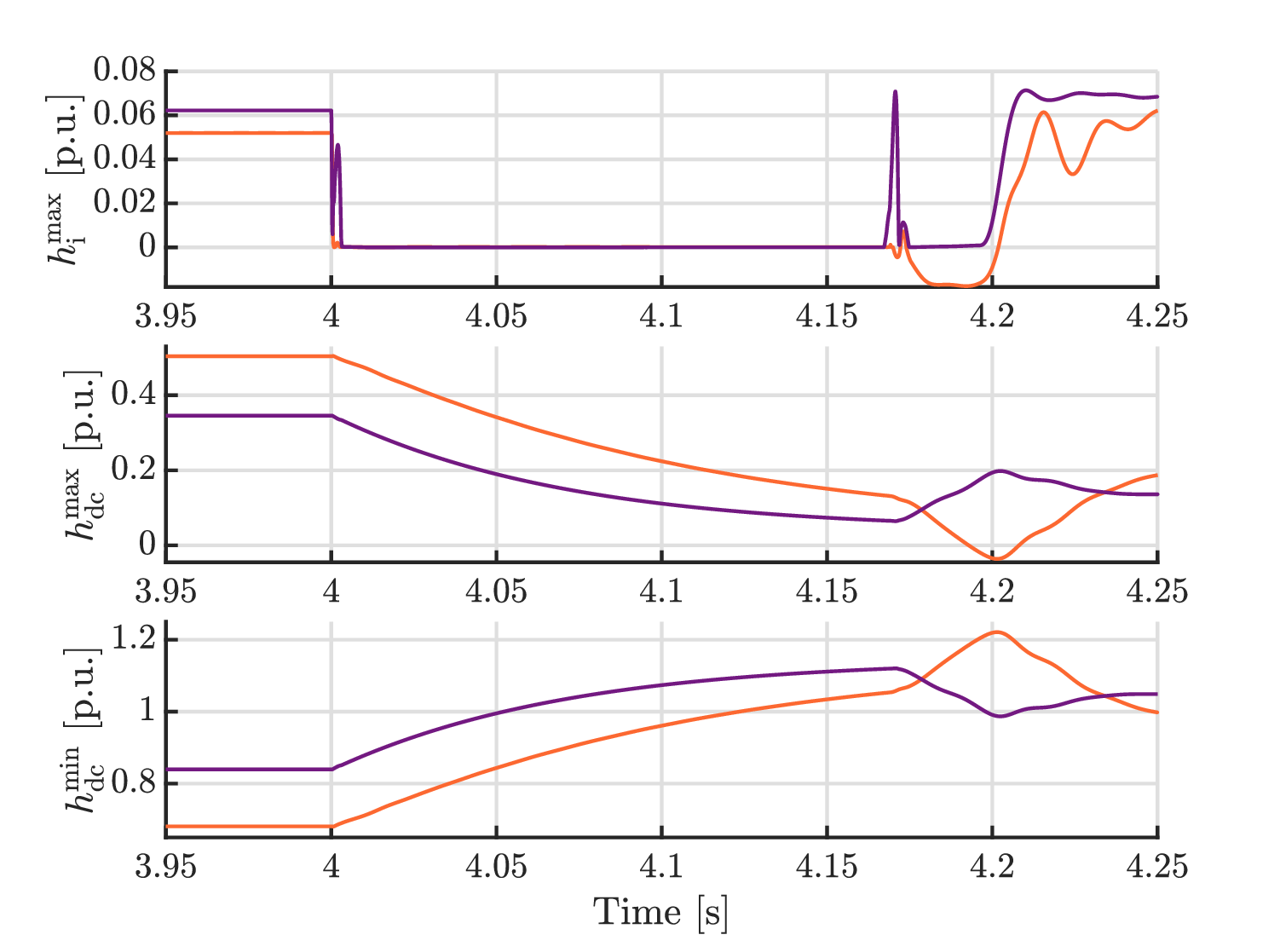}}
  \caption{CBFs encoding current and dc voltage limits for $\text{PV}_{\mathrm{1}}$ (\textcolor[rgb]{0.9882, 0.4078, 0.1922}{\rule[0.5ex]{1.2em}{0.6pt}}) and  $\text{PV}_{\mathrm{2}}$ (\textcolor[rgb]{0.4510, 0.0980, 0.5059}{\rule[0.5ex]{1.2em}{0.6pt}}) during the short-circuit fault.}
  \label{fig.hFault}
  \end{figure}

Finally, the capability of the proposed GFM control to withstand symmetric short-circuit faults is evaluated. To this end, we consider a separate scenario in which $\mathrm{s}_3$ is closed from $t=4$~s to $t=4.166$~s to simulate a $10$-cycle three-phase short circuit. The simulation results for this scenario are illustrated in Figs.~\ref{fig.abcFault},~\ref{fig.viFault},~\ref{fig.PQFault}, and~\ref{fig.hFault}.

Due to the short-circuit fault, the output voltages, particularly that of $\text{PV}_{\mathrm{2}}$, drop to near zero, while the VSC ac currents reach their limits. The power injections of the GFM PV units also approach zero, causing the dc voltages to approach their open-circuit values and PV power generation to cease. The average frequencies of the GFM PV units, obtained from the time derivatives of $\theta_{\mathrm{1}}(t)$ and $\theta_{\mathrm{2}}(t)$ during the fault, are $\omega_{\mathrm{1}} = 1.01$~p.u. and $\omega_{\mathrm{2}} = 1.06$~p.u., respectively.

It can be observed that the CBFs remain nonnegative throughout the fault, i.e., the  GFM PV units remain within their constraints. Following fault clearance, the GFM controls resynchronize, and the associated phase jumps result in brief, small transient excursions beyond the dc-bus voltage and long-term current limits. These small transient excursions are within an acceptable range for both the current and dc voltage limits. Subsequently, the constrained GFM control returns the dc-bus voltages and output currents of the single-stage PV systems to within their safe bounds without inducing severe transients.

\section{Comparison with Representative GFM Control Strategies}
Finally, we use the single-stage GFM PV unit connected to an infinite bus in Fig.~\ref{fig.PVsingleBus} to compare the performance of the proposed constrained GFM control with that of two benchmark strategies that incorporate widely studied current, power, and dc voltage limiters. The PV array operates at an irradiance of $1000$~$\mathrm{W/m^2}$, and the dc voltage setpoint is $v_{\mathrm{dc}}^{\star}=1.14$~p.u. For this comparison, we select $R_{\mathrm{d}}=0.17$~p.u., $\tau_{\mathrm{v}}=14$~ms, $\gamma_{\theta}=0.08$~p.u., $n_{\mathrm{itr}}=10$, $\sigma=2$, and $\rho=5$. The remaining parameters of the system are identical to the multi-source system parameters summarized in Table~\ref{tab.Multiparameters}. The system is again modeled in MATLAB/Simulink using the Specialized Power Systems library, with a simulation time step of $10$~$\mu$s.

To the authors' knowledge, no GFM control has been reported in the literature that explicitly accounts for all constraints considered in this work. Thus, the proposed method is evaluated against two benchmark controls that combine well-studied current, power, and dc-voltage limiters that were designed separately. In particular, we apply the PI-based dc-voltage limiter introduced in~\cite{chen2019power} to enforce both the lower and upper dc-voltage limits. In parallel, the power-limiting droop control in~\cite{iraniparast2025networked} is implemented to limit the injected power to the MPP power. Additionally, the hybrid threshold virtual impedance (HTVI) in~\cite{zeng2023hybrid} and the variable virtual impedance (VVI) in~\cite{qoria2022variable} are implemented alongside the dc-voltage and power-limiting loops to limit the converter current. The resulting benchmark controls enforce each constraint through a separate, established limiting scheme.

\begin{figure}
  	\centerline{\includegraphics[scale=0.39,trim={52cm 0.6cm 53cm 0.5cm}]{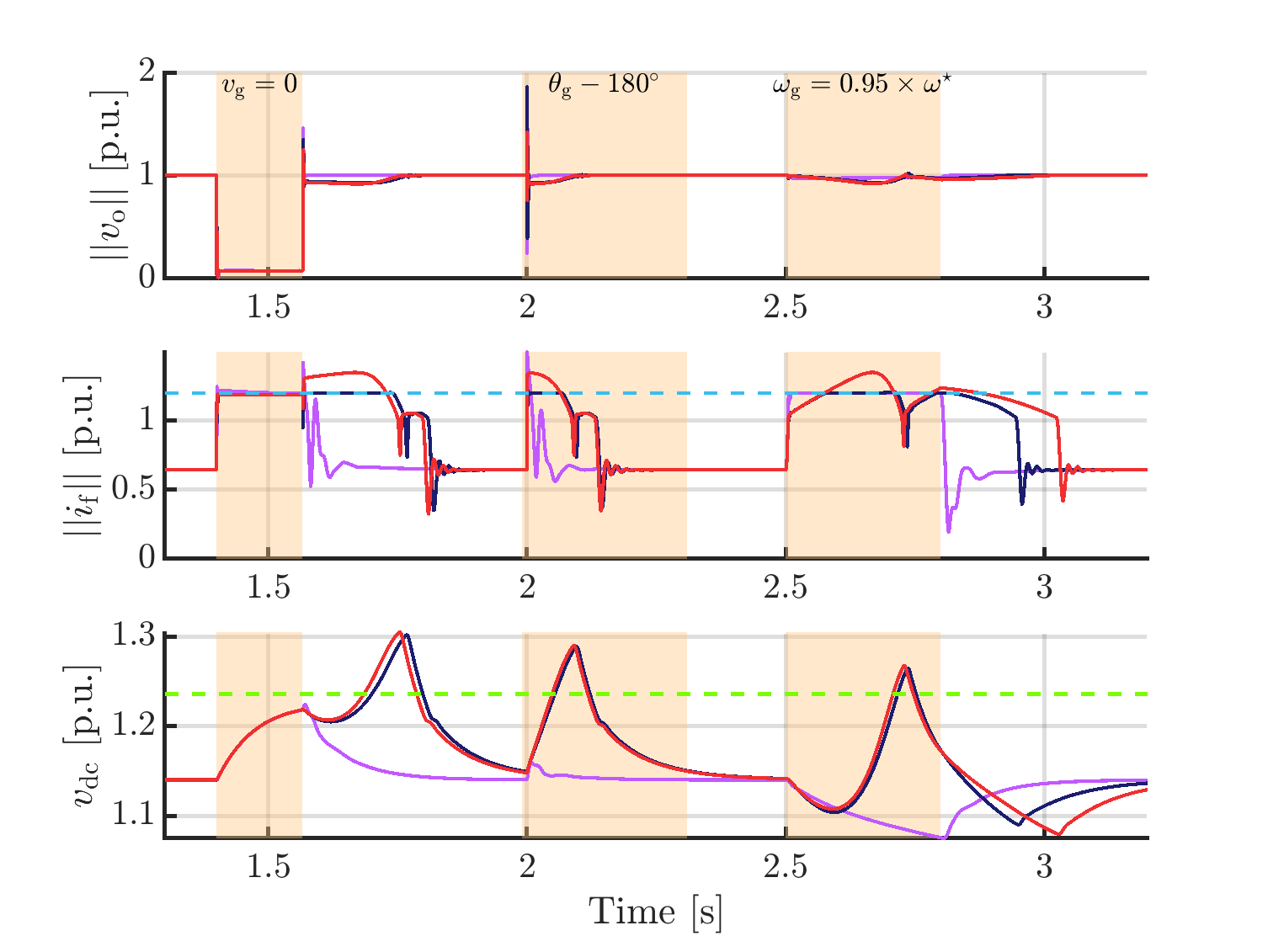}}
  	\caption{Voltage magnitude, current magnitude, and dc voltage for the proposed control  (\textcolor[rgb]{0.7569, 0.3490, 1.0000}{\rule[0.5ex]{1.2em}{0.6pt}}), the HTVI method  (\textcolor[rgb]{0.1059, 0.1137, 0.4275}{\rule[0.5ex]{1.2em}{0.6pt}}), and the $\text{VVI}$ method (\textcolor[rgb]{0.9412, 0.1843, 0.1882}{\rule[0.5ex]{1.2em}{0.6pt}}) as well as maximum dc voltage (\textcolor[rgb]{0.48627, 0.98824, 0.0}{\rule[0.5ex]{0.32em}{0.6pt}\hspace{0.12em}\rule[0.5ex]{0.32em}{0.6pt}\hspace{0.12em}\rule[0.5ex]{0.32em}{0.6pt}}) and maximum current (\textcolor[rgb]{0.2235, 0.7412, 0.9294}{\rule[0.5ex]{0.32em}{0.6pt}\hspace{0.12em}\rule[0.5ex]{0.32em}{0.6pt}\hspace{0.12em}\rule[0.5ex]{0.32em}{0.6pt}}).}
  	\label{fig.VIcomp}
  \end{figure}

\begin{figure}
  	\centerline{\includegraphics[scale=0.39,trim={52cm 0.6cm 53cm 0.5cm}]{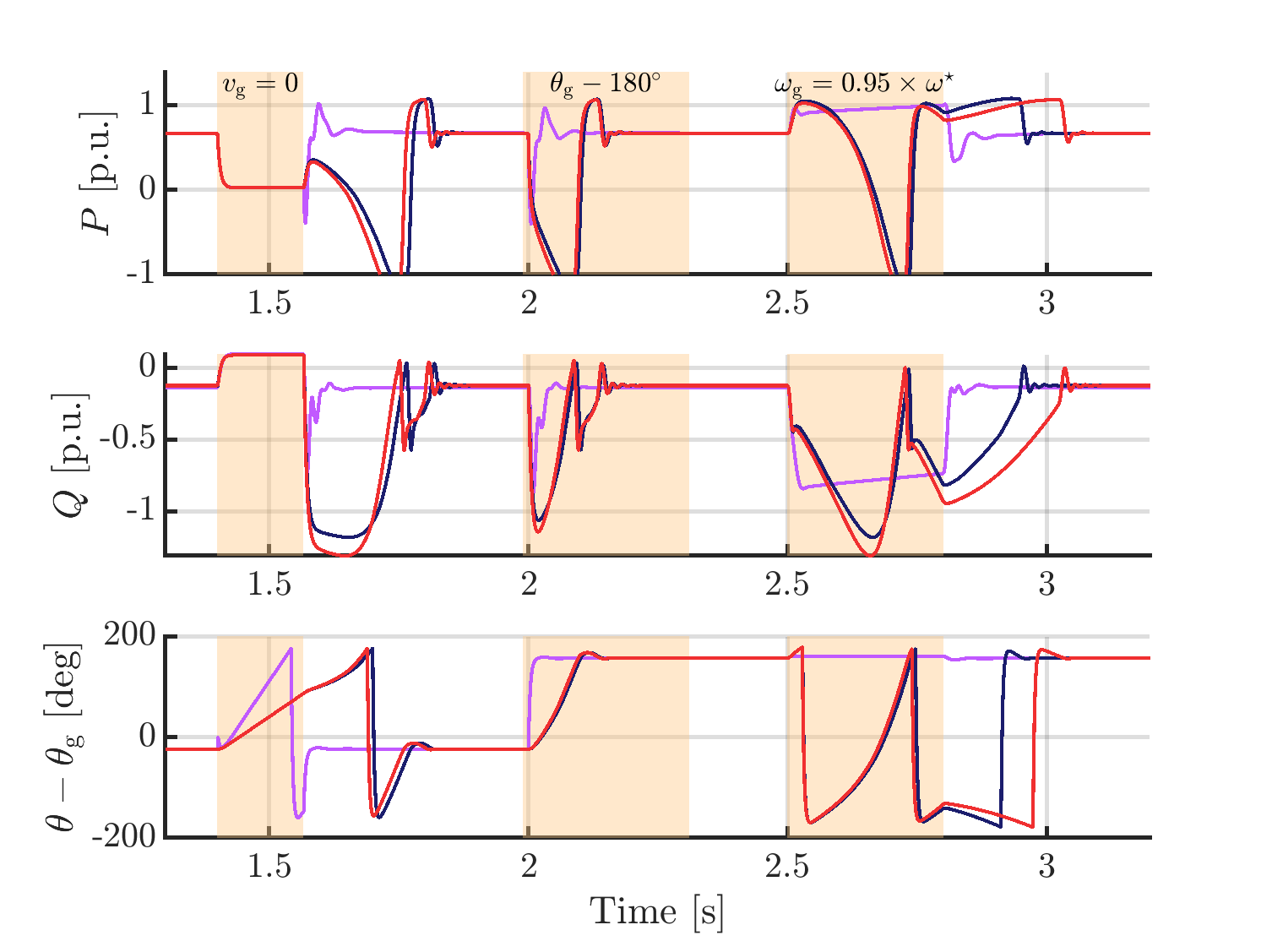}}
  	\caption{Active powers, reactive powers, and phase-angle differences for the proposed method  (\textcolor[rgb]{0.7569, 0.3490, 1.0000}{\rule[0.5ex]{1.2em}{0.6pt}}), the HTVI method  (\textcolor[rgb]{0.1059, 0.1137, 0.4275}{\rule[0.5ex]{1.2em}{0.6pt}}), and the $\text{VVI}$ method (\textcolor[rgb]{0.9412, 0.1843, 0.1882}{\rule[0.5ex]{1.2em}{0.6pt}}).}
  	\label{fig.PQcomp}
  \end{figure}

Simulation results for three scenarios are shown in Figs.~\ref{fig.VIcomp} and~\ref{fig.PQcomp}. First, a three-phase short-circuit fault is applied at $t=1.4$~s by setting the infinite bus voltage magnitude $V_{\mathrm{g}}$ to zero. The fault lasts for $10$ cycles and is cleared at $t=1.56$~s. The ac current remains within its limit using HTVI control but exceeds its threshold using VVI control. Moreover, the dc voltage exceeds its maximum allowable value using both controls, and the single-stage PV system exhibits prolonged post-fault recovery. In contrast, the proposed constrained GFM control handles the fault effectively and rapidly recovers following a small and brief excursion past the long-term current limit.

The next scenario considers a $-180^{\circ}$ phase jump of the grid voltage at $t=2$~s. This contingency evaluates the resynchronization capability of the system from an initial condition with the worst-case phase-angle difference (i.e., after a loss of synchronization). The benchmark controls fail to maintain the dc voltage within its safe range and respond slowly to the phase jump, without settling within an acceptable time. In contrast, the proposed constrained GFM control exhibits a satisfactory response with no significant current-limit violation.

Finally, the grid frequency is reduced by $5\%$ to investigate the low-frequency ride-through capability of both control strategies. To this end, we set $\omega_{\mathrm{g}} = 0.95 \times \omega^{\star}$ at $t=2.4$~s and then return the grid frequency to its nominal value at $t=2.7$~s. During this interval, the benchmark controls lose synchronization and fail to inject the active power expected for primary frequency control. As in the previous scenarios, the VVI control also fails to maintain the current within its prescribed limits. Moreover, the dc voltage exceeds its maximum allowable value under both methods, and neither control achieves sufficiently fast post-contingency recovery of the PV system. In contrast, the proposed constrained control remains synchronized, adjusts its frequency to match the grid frequency, and continues supporting the grid up to its available capacity without violating the physical limits.

It is worth noting that neither the combination of the dc voltage limiter in \cite{chen2019power} and the power-limiting droop control nor the HTVI and VVI controls can independently handle faults and large disturbances without leading to dc-voltage collapse or current-limit violations. Even when these limiting methods are combined, as in the benchmark strategies considered here, the use of multiple parallel control loops introduces additional coordination and tuning complexity. Furthermore, incorporating additional physical constraints generally requires adding new limiting loops, which reduces scalability and may introduce undesirable interactions with the existing control loops. Such interactions and lack of coordination can degrade the overall transient performance and, under severe operating conditions, may compromise system stability.

\section{Conclusion}
In this work, we developed a novel constrained dual-port GFM controller for single-stage PV systems. The proposed control systematically accounts for key physical and operational constraints, including constraints on ac current, dc-bus voltage, and converter modulation. These constraints are encoded by individual scalar CBFs, whose non-negativity characterizes safe operation of the GFM unit. The single-stage PV dynamics are then used to derive the set of admissible converter ac voltages that ensure CBF non-negativity over time. The proposed controller retains the features of unconstrained dual-port GFM control to the extent possible while satisfying the PV system constraints. To this end, the controller selects the admissible converter ac voltage that minimizes the deviation from the phase-angle and magnitude references generated by the unconstrained GFM controller. The deviation from the unconstrained GFM control is measured using a cost function that provides flexibility in shaping the phase-angle/frequency and magnitude response during faults, overloads, and other events that activate converter limits. An efficient algorithm is developed for real-time implementation. The effectiveness and performance of the resulting constrained GFM control are evaluated through EMT simulations in MATLAB/Simulink. Further analysis of the proposed CBF-based constrained GFM control, particularly its stability properties, is an interesting direction for future research.

\bibliographystyle{IEEEtran}
\bibliography{IEEEabrv,ref}

\end{document}